\documentclass[10pt,conference,letterpaper]{IEEEtran}

\usepackage[acronyms,nonumberlist,nopostdot,nomain,nogroupskip]{glossaries}

\AtBeginDocument{%
  \providecommand\BibTeX{{%
    \normalfont B\kern-0.5em{\scshape i\kern-0.25em b}\kern-0.8em\TeX}}}
 \IEEEoverridecommandlockouts

\usepackage{amsmath}
\usepackage{mathabx}
\usepackage{bbold}
\usepackage{array}
\usepackage{bm}
\usepackage{colortbl}
\usepackage{comment}
\usepackage{enumitem}
\usepackage{epsfig}
\usepackage[para]{footmisc}
\usepackage{graphicx}
\usepackage{wrapfig}
\usepackage[utf8]{inputenc}
\usepackage{multirow}
\usepackage{microtype}
\usepackage{subcaption}
\usepackage{tablefootnote}
\usepackage{tabularx}
\usepackage{url}
\usepackage{xcolor}
\usepackage{url}
\usepackage[numbers,sort&compress]{natbib}

\usepackage{xspace}
\newcommand{\KU}{\texttt{Katch-Up}\xspace}
\newcommand{\FW}{\texttt{SmartDet}\xspace}
\newacronym{6g}{6G}{sixth generation}
\newacronym{3gpp}{3GPP}{3rd Generation Partnership Project}
\newacronym{adc}{ADC}{Analog to Digital Converter}
\newacronym{dac}{DAC}{Digital to Analog Converter}
\newacronym{5g}{5G}{5th generation}
\newacronym{aimd}{AIMD}{Additive Increase Multiplicative Decrease}
\newacronym{am}{AM}{Acknowledged Mode}
\newacronym{amc}{AMC}{Adaptive Modulation and Coding}
\newacronym{aoa}{AoA}{Angle of Arrival}
\newacronym{aod}{AoD}{Angle of Departure}
\newacronym{aqm}{AQM}{Active Queue Management}
\newacronym{awgn}{AGWN}{Additive White Gaussian Noise}
\newacronym{balia}{BALIA}{Balanced Link Adaptation}
\newacronym{bdp}{BDP}{Bandwidth-Delay Product}
\newacronym{bf}{BF}{Beamforming}
\newacronym{fpga}{FPGA}{field-programmable gate array}
\newacronym{cc}{CC}{Congestion Control}
\newacronym{cdf}{CDF}{Cumulative Distribution Function}
\newacronym{cn}{CN}{Core Network}
\newacronym{cm}{CM}{confusion matrix}
\newacronym[plural=\gls{cnn}s,firstplural=convolutional neural networks (CNNs)]{cnn}{CNN}{convolutional neural network}
\newacronym{cqi}{CQI}{Channel Quality Information}
\newacronym{cp}{CP}{Control Plane}
\newacronym{csirs}{CSI-RS}{Channel State Information - Reference Signal}
\newacronym{dc}{DC}{Dual Connectivity}
\newacronym{dce}{DCE}{Direct Code Execution}
\newacronym{dci}{DCI}{Downlink Control Information}
\newacronym{dmr}{DMR}{Deadline Miss Ratio}
\newacronym{drl}{DRL}{Deep Reinforcement Learning}
\newacronym{dmrs}{DMRS}{DeModulation Reference Signal}
\newacronym{e2e}{E2E}{End-to-End}
\newacronym{ecn}{ECN}{Explicit Congestion Notification}
\newacronym{ebs}{EBS}{exhaustive beam sweep}
\newacronym{edf}{EDF}{Earliest Deadline First}
\newacronym{enb}{eNB}{evolved Node Base}
\newacronym{epc}{EPC}{Evolved Packet Core}
\newacronym{es}{ES}{Edge Server}
\newacronym{fps}{fps}{frames per second}
\newacronym{fdma}{FDMA}{Frequency Division Multiple Access}
\newacronym{fdd}{FDD}{Frequency Division Duplexing}
\newacronym[firstplural=Radio Access Technologies (RATs)]{rat}{RAT}{Radio Access Technology}
\newacronym{fs}{FS}{Fast Switching}
\newacronym{txer}{TX}{transmitter}
\newacronym{rxer}{RX}{receiver}
\newacronym{bt}{BT}{beam tracking}
\newacronym{ftp}{FTP}{File Transfer Protocol}
\newacronym{gnb}{gNB}{Next Generation Node Base}
\newacronym{bs}{BS}{Base Station}
\newacronym{harq}{HARQ}{Hybrid Automatic Repeat reQuest}
\newacronym{hetnet}{HetNet}{Heterogeneous Network}
\newacronym{hh}{HH}{Hard Handover}
\newacronym{hol}{HOL}{Head-of-Line}
\newacronym{ia}{IA}{initial access}
\newacronym{imt}{IMT}{International Mobile Telecommunication}
\newacronym{iot}{IoT}{Internet of Things}
\newacronym{iou}{IoU}{Intersection over Union}
\newacronym{los}{LOS}{Line-of-Sight}
\newacronym{lte}{LTE}{Long Term Evolution}
\newacronym{m2m}{M2M}{Machine to Machine}
\newacronym{ml}{ML}{machine learning}
\newacronym{dl}{DL}{deep learning}
\newacronym{mac}{MAC}{Medium Access Control}
\newacronym{map}{mAP}{mean Average Precision}
\newacronym{mar}{mAR}{mean Average Recall}
\newacronym{mc}{MC}{Multi-Connectivity}
\newacronym{mcs}{MCS}{Modulation and Coding Scheme}
\newacronym{mec}{MEC}{Mobile Edge Cloud}
\newacronym{mi}{MI}{Mutual Information}
\newacronym{mimo}{MIMO}{Multiple Input, Multiple Output}
\newacronym{mmwave}{mmWave}{millimeter wave}
\newacronym{mmWave}{mmWave}{Millimeter wave}
\newacronym{mptcp}{MPTCP}{Multipath TCP}
\newacronym{mr}{MR}{Maximum Rate}
\newacronym{mss}{MSS}{Maximum Segment Size}
\newacronym{mtd}{MTD}{Machine-Type Device}
\newacronym{mtu}{MTU}{Maximum Transmission Unit}
\newacronym{nfv}{NFV}{Network Function Virtualization}
\newacronym{nlos}{NLOS}{Non-Line-of-Sight}
\newacronym{nr}{NR}{New Radio}
\newacronym{ofdm}{OFDM}{Orthogonal Frequency Division Multiplexing}
\newacronym{pdcch}{PDCCH}{Physical Downlink Control Channel}
\newacronym{pdcp}{PDCP}{Packet Data Convergence Protocol}
\newacronym{pdsch}{PDSCH}{Physical Downlink Shared Channel}
\newacronym{pdu}{PDU}{Packet Data Unit}
\newacronym{pf}{PF}{Proportional Fair}
\newacronym{pgw}{PGW}{Packet Gateway}
\newacronym{phy}{PHY}{Physical}
\newacronym{pbch}{PBCH}{Physical Broadcast Channel}
\newacronym[plural=\gls{mme}s,firstplural=Mobility Management Entities (MMEs)]{mme}{MME}{Mobility Management Entity}
\newacronym{prb}{PRB}{Physical Resource Block}
\newacronym{pss}{PSS}{Primary Synchronization Signal}
\newacronym{pucch}{PUCCH}{Physical Uplink Control Channel}
\newacronym{pusch}{PUSCH}{Physical Uplink Shared Channel}
\newacronym{rach}{RACH}{Random Access Channel}
\newacronym{ran}{RAN}{Radio Access Network}
\newacronym{red}{RED}{Random Early Detection}
\newacronym{rf}{RF}{Radio Frequency}
\newacronym{rlc}{RLC}{Radio Link Control}
\newacronym{rlf}{RLF}{Radio Link Failure}
\newacronym{rrc}{RRC}{Radio Resource Control}
\newacronym{rrm}{RRM}{Radio Resource Management}
\newacronym{rr}{RR}{Round Robin}
\newacronym{rs}{RS}{Remote Server}
\newacronym{rsrp}{RSRP}{Reference Signal Received Power}
\newacronym{rss}{RSS}{Received Signal Strength}
\newacronym{rtt}{RTT}{Round Trip Time}
\newacronym{rw}{RW}{Receive Window}
\newacronym{rx}{RX}{Receiver}
\newacronym{sa}{SA}{standalone}
\newacronym{sack}{SACK}{Selective Acknowledgment}
\newacronym{sap}{SAP}{Service Access Point}
\newacronym{ap}{AP}{Access Point}
\newacronym{sch}{SCH}{Secondary Cell Handover}
\newacronym{scoot}{SCOOT}{Split Cycle Offset Optimization Technique}
\newacronym{sdma}{SDMA}{Spatial Division Multiple Access}
\newacronym{sinr}{SINR}{Signal to Interference plus Noise Ratio}
\newacronym{sm}{SM}{Saturation Mode}
\newacronym{snr}{SNR}{Signal-to-Noise-Ratio}
\newacronym{son}{SON}{Self-Organizing Network}
\newacronym{ss}{SS}{Synchronization Signal}
\newacronym{ssbs}{SSBs}{synchronization signal blocks}
\newacronym{ssb}{SSB}{synchronization signal block}
\newacronym{srs}{SRS}{Sounding Reference Signal}
\newacronym{sss}{SSS}{Secondary Synchronization Signal}
\newacronym{tb}{TB}{Transport Block}
\newacronym{tcp}{TCP}{Transmission Control Protocol}
\newacronym{tdd}{TDD}{Time Division Duplexing}
\newacronym{tdma}{TDMA}{Time Division Multiple Access}
\newacronym{tfl}{TfL}{Transport for London}
\newacronym{tm}{TM}{Transparent Mode}
\newacronym{trp}{TRP}{Transmitter Receiver Pair}
\newacronym{tti}{TTI}{Transmission Time Interval}
\newacronym{ttt}{TTT}{Time-to-Trigger}
\newacronym{tx}{TX}{Transmitter}
\newacronym{ue}{UE}{User Equipment}
\newacronym{ul}{UL}{Uplink}
\newacronym{uml}{UML}{Unified Modeling Language}
\newacronym{um}{UM}{Unacknowledged Mode}
\newacronym{utc}{UTC}{Urban Traffic Control}
\newacronym{vm}{VM}{Virtual Machine}
\newacronym{rsrq}{RSRQ}{Reference Signal Received Quality}
\newacronym{rssi}{RSSI}{Received Signal Strength Indicator}
\newacronym{crs}{CRS}{Cell Reference Signal}
\newacronym{nsa}{NSA}{Non Stand Alone}
\newacronym{mrdc}{MR-DC}{Multi \gls{rat} \gls{dc}}
\newacronym{endc}{EN-DC}{E-UTRAN-\gls{nr} \gls{dc}}
\newacronym{5gc}{5GC}{5G Core}
\newacronym{si}{SI}{Study Item}
\newacronym{iab}{IAB}{Integrated Access and Backhaul}
\newacronym{wf}{WF}{Wired-first}
\newacronym{hqf}{HQF}{Highest-quality-first}
\newacronym{pa}{PA}{Position-aware}
\newacronym{mlr}{MLR}{Maximum-local-rate}
\newacronym{wbf}{WBF}{Wired Bias Function}
\newacronym{mib}{MIB}{Master Information Block}
\newacronym{sib}{SIB}{Secondary Information Block}
\newacronym{kpi}{KPI}{Key Performance Indicator}
\newacronym{ppp}{PPP}{Poisson Point Process}
\newacronym{gtp}{GTP}{GPRS Tunneling Protocol}
\newacronym{amf}{AMF}{Access and Mobility Management Function}
\newacronym{dash}{DASH}{Dynamic Adaptive Streaming over HTTP}
\newacronym{http}{HTTP}{HyperText Transfer Protocol}
\newacronym{qos}{QoS}{Quality of Service}
\newacronym{udp}{UDP}{User Datagram Protocol}
\newacronym{cu}{CU}{Central Unit}
\newacronym{du}{DU}{Distributed Unit}
\newacronym{mt}{MT}{Mobile Termination}
\newacronym{sdap}{SDAP}{Service Data Adaptation Protocol}
\newacronym{tdm}{TDM}{Time Division Multiplexing}
\newacronym{fdm}{FDM}{Frequency Division Multiplexing}
\newacronym{sdm}{SDM}{Space Division Multiplexing}
\newacronym{dag}{DAG}{Directed Acyclic Graph}
\newacronym{st}{ST}{Spanning Tree}
\newacronym{ummimo}{UM-MIMO}{Ultra-massive Multiple Input, Multiple Output}
\newacronym{uavs}{UAVs}{Unmanned Aerial Vehicles}
\newacronym{wlan}{WLAN}{Wireless LAN}
\newacronym{rlnc}{RLNC}{Random Linear Network Coding}
\newacronym{drx}{DRX}{Discontinuous Reception}
\newacronym{cpu}{CPU}{Central Processing Unit}
\newacronym{txb}{TXB}{transmitter's beam}
\newacronym{rxb}{RXB}{receiver's beam}
\newacronym{sifs}{SIFS}{Short Interframe Space}
\newacronym{difs}{DIFS}{DCF Interframe Space}

\newacronym{rfid}{RFID}{Radio Frequency Identification}
\newacronym{rfp}{RFP}{radio fingerprinting}
\newacronym{sdr}{SDR}{software-defined radio}
\newacronym{dnn}{DNN}{deep neural network}

\newacronym{od}{OD}{object detection}

\newacronym{ot}{OT}{object tracking}

\begin{document}

\title{\FW: Context-Aware Dynamic Control of Edge Task Offloading for Mobile Object Detection\vspace{-0.2cm}}

\author{Davide Callegaro$^*$, Francesco Restuccia$^\dagger$ and Marco Levorato$^*$\\
$*$ Computer Science Dept., University of California, Irvine, United States\\
$\dagger$ Department of Electrical and Computer Engineering, Northeastern University, United States \vspace{-0.2cm}
\\
\thanks{This work was supported by the Intel Corporation and the NSF grant MLWiNS-2003237.}
}




\maketitle
\pagestyle{plain}

\glsresetall

\begin{abstract}
Mobile devices increasingly rely on object detection (OD) through deep neural networks (DNNs) to perform critical tasks. Due to their high complexity, the execution of these DNNs requires excessive time and energy. Low-complexity object tracking (OT) can be used with OD, where the latter is periodically applied to generate “fresh” references for tracking. However, the frames processed with OD incur large delays, which may make the reference outdated and degrade tracking quality. Herein, we propose to use edge computing in this context, and establish \textit{parallel} OT (at the mobile device) and OD (at the edge server) processes that are resilient to large OD latency. We propose \KU, a novel tracking mechanism that improves the system resilience to excessive OD delay. However, while \KU significantly improves performance, it also increases the computing load of the mobile device. Hence, we design \FW, a low-complexity controller based on deep reinforcement learning (DRL) that learns controlling the trade-off between resource utilization and OD performance. \FW takes as input context-related information related to the current video content and the current network conditions to optimize frequency and type of OD offloading, as well as \KU utilization. We extensively evaluate \FW on a real-world testbed composed of a JetSon Nano as mobile device and a GTX 980 Ti as edge server, connected through a Wi-Fi link. Experimental results show that \FW achieves an optimal balance between tracking performance – mean Average Recall (mAR) and resource usage. With respect to a baseline with full \KU usage and maximum channel usage, we still increase mAR by 4\% while using 50\% less of the channel and 30\% power resources associated with \KU. With respect to a fixed strategy using minimal resources, we increase mAR by 20\% while using \KU on 1/3 of the frames.

\end{abstract}

\section{Introduction}

\begin{figure}[!h]
    \centering
    \includegraphics[width=0.9\columnwidth]{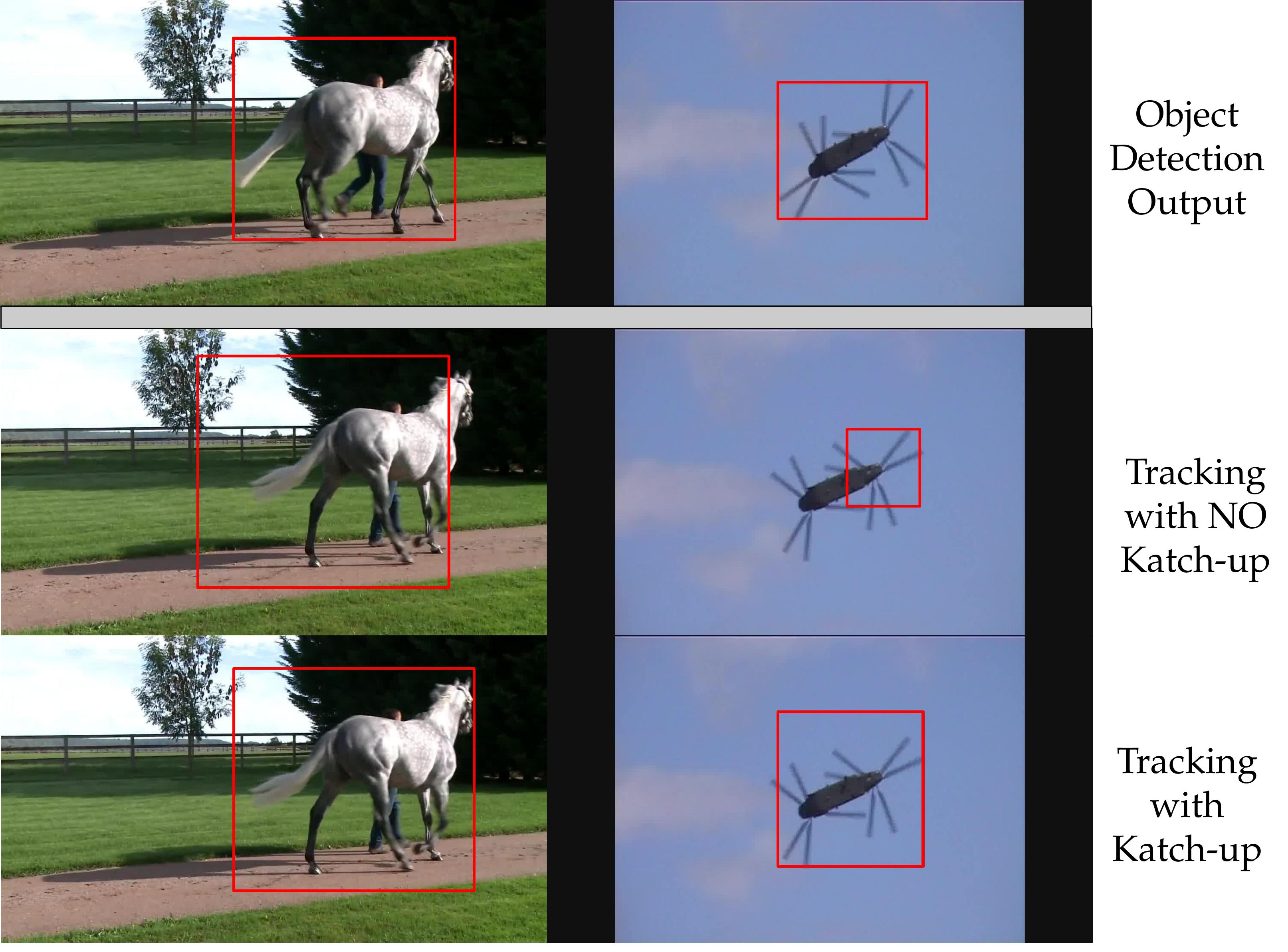}
    \caption{Examples of bounding boxes produced by tracking with and without \KU ($400$ms object detection delay).\vspace{-0.4cm}}
    \label{fig:bboxes}
\end{figure}

Real-time \gls{od} is a critical component of a wide array of current and future applications and systems, including autonomous vehicles \cite{50125} and city-monitoring \cite{distream}. In a nutshell, \gls{od} aims at the precise identification and positioning of objects contained in an image or a sequence of images. The outcome is a set of bounding boxes (see Fig.~\ref{fig:bboxes} for a graphical example) and associated labels describing the objects.

The majority of existing frameworks leverages \glspl{dnn} to perform \gls{od} \cite{9155438, internet_battlefield}. However, state-of-the-art \glspl{dnn} have very large complexity and cannot be entirely executed on mobile devices \cite{benchmark_od}. While lower-complexity algorithms exist \cite{yao2017deepiot, deepadapter}, they achieve poor performance -- \emph{e.g.}, measured as accuracy, recall, or precision (see Section~\ref{sec:background} for a definition of the metrics) -- compared to state-of-the-art models. For instance, Yolo-Lite \cite{yolov3} achieves a frame rate of 22 frames per second on embedded devices, but has a mean average precision (mAP) of 12.36\% on the COCO dataset \cite{coco_dataset}. 
EfficientDet 0-7 \cite{tan2020efficientdet} is a family of state-of-the-art \gls{od} models that offer increasing performance at the price of an increasing complexity. EfficientDet 7 achieves \gls{map} of 55.1\%, but leverages 52M parameters. Even EfficientDet 0, the simplest model in the family, which achieves 33\% on the COCO dataset, is 2x times more complex than SSD-MobileNet v2: a lower-performance DNN specifically designed for mobile platforms,
which achieves \gls{map} of 20\%, and in our experiments can provide up 6 \gls{fps}, while significantly increasing power consumption. We remark that pruning and quantization, two techniques widely used to make DNN simplers, greatly degrade OD performance. 

\begin{figure}[h]
    \centering
    \includegraphics[width=\columnwidth]{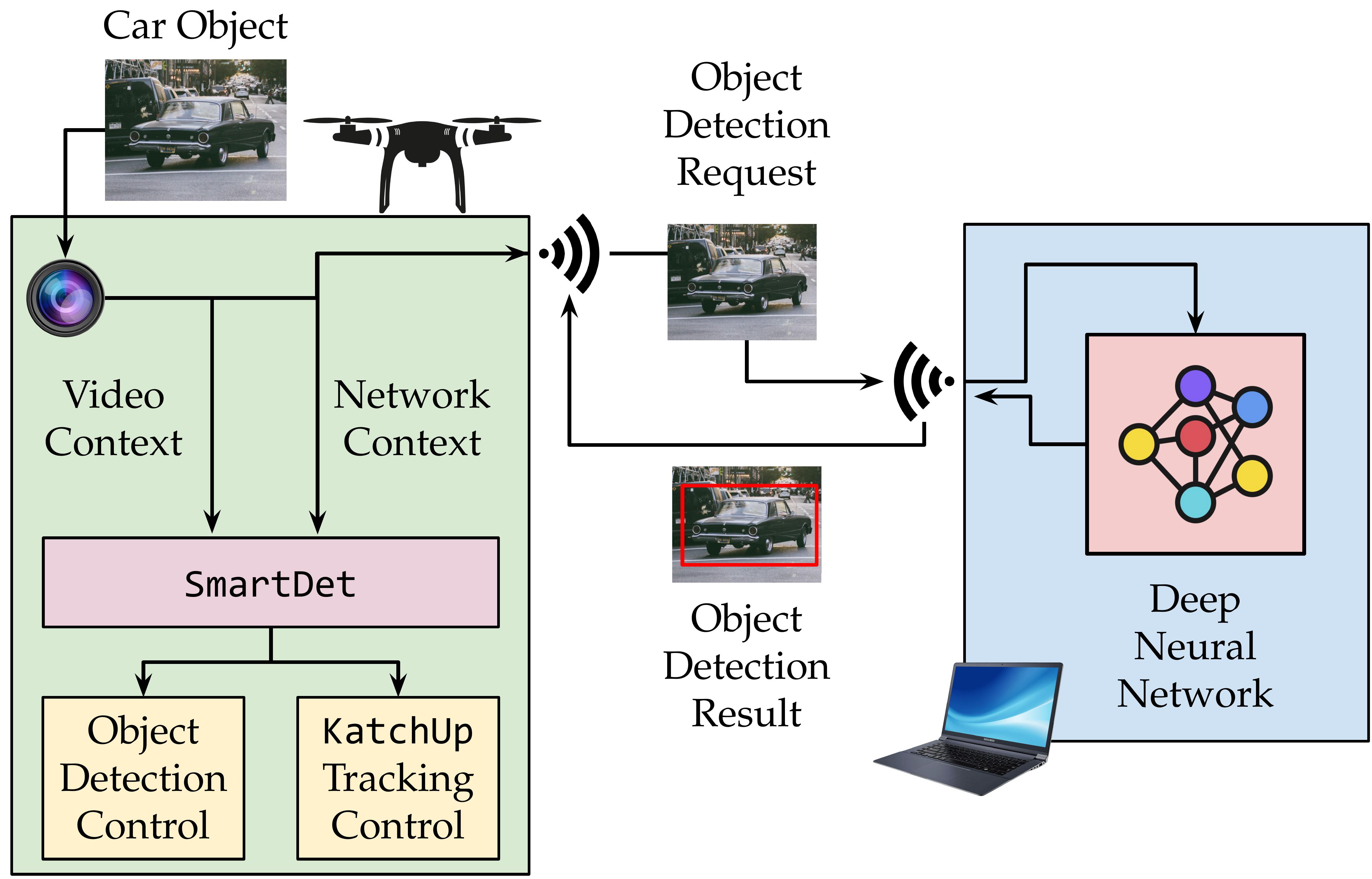}
    \caption{Overview of \FW main components.}
    \label{fig:framework}
\end{figure}

There are two main strategies to address this issue: (1) \emph{Edge Computing:}~\cite{liu2019edge} the mobile device offloads the OD stream to edge servers - compute capable devices located at the wireless network's edge; (2) \emph{Object Tracking:} using a lower complexity object tracker in conjunction with a high(er) complexity object detector. Prior work considers these two strategies in isolation, and we contend that such approach fails to provide acceptable performance in many settings of great relevance (see discussion below). Our paper presents \FW: the first framework to propose the use of object detection and tracking in an edge computing setting. Notably, the composition of the two strategies presents both unique challenges and opportunities, which we are the first to explore in this paper.

Before introducing \FW, we first discuss the two strategies mentioned above.

\noindent
(1) \emph{Joint Detection and Tracking:} To address the excessive computational overhead associated with \gls{od}, \gls{ot} is often used in mobile computing contexts~\cite{8653851}. Trackers assume temporal correlation in the sequence of images, and use a previously computed reference to analyze a new image~\cite{xu2021approxnet}. 
The idea behind \gls{ot} is fairly simple; given a video, \gls{od} is performed periodically, and its outcome serves as reference for \gls{ot} on the remaining frames. 
Since \gls{ot} is less computationally expensive than \gls{od}, energy consumption and computing load are reduced \cite{detect_and_track}.
However, due to constraints in the computing power of mobile devices, the execution of \gls{od} may take a large amount of time. We show in Section~\ref{sec:results1} that an outdated \gls{od} reference can degrade \gls{mar} performance by up to 25\% on the average due to these effects, which are exacerbated in videos with highly dynamic objects.

(2) \emph{Edge Computing:} prior work considered approaches where all the frames are processed using gls{od}, and mobile devices offload the streams of \gls{od} tasks to edge servers. This partially addresses the issue of high computational complexity, as edge servers has considerably more computing power and energy resources compared to mobile devices. However, an approach purely based on offloading \gls{od} has the following drawbacks: (\emph{a}) wireless channels have usually a constrained and erratic capacity, especially in applications such as autonomous vehicles where mobile devices are often moving. This leads to high communication latency and large latency variations \cite{callegaro2021seremas, benchmark_od}; (\emph{b}) frequent transfer of images consumes a large amount of channel capacity -- \emph{e.g.}, up to 20\% of available Wi-Fi bandwidth in our experiments -- possibly resulting in channel congestion \cite{wang2020joint}; (\emph{c}) as all frames are transferred to the edge server for analysis, each mobile device imposes a considerable processing load to the edge server. 
Existing work focuses on scenarios where the wireless link capacity is extremely large and substantially steady (\emph{e.g.}, $350$Mbps, and the edge servers have -- individually or collectively -- high computing power (\emph{e.g.}, see~\cite{elf}).

In contrast with existing work, in this paper we address the challenging scenarios where \textit{the capacity of the wireless channel is limited and erratic, and the edge servers have limited computing power}. We propose to establish two parallel processes: the mobile device executes \gls{ot} on all frames, and only some of the frames are sent to the edge server for \gls{od}. This approach assigns to the mobile device a lightweight analysis process, thus reducing the requirements on available resources and takes advantage of the greater computing power of edge servers, while imposing a moderate communication and computing load. However, in order to maximize the performance of such system there are several challenges that need to be addressed: (\emph{1}) variations in the capacity of the channel may still result in some of the \gls{od} references to refer to outdated frames, which may harm tracking performance; (\emph{2}) the tracking performance is greatly influenced not only by reference delay, but by other parameters such as \gls{od} period and accuracy -- which in turn determine channel and server load. To address the above key issues, this paper makes the following novel contributions:\smallskip

$\bullet$ We introduce a new tracking strategy, which we refer to as \KU (Section~\ref{sec:KU}), to make the edge-mobile system resilient to \gls{od} delay. In \KU, when an object detection outcome is received, we re-track the sequence of images starting from the time at which the frame was generated. This technique greatly improves the quality of the reference available to tracking against \gls{od} delay, thus boosting performance. Fig.~\ref{fig:bboxes} shows two examples of pictures where \KU was and was not applied to the tracking process - the better quality of the bounding boxes is apparent;\smallskip

$\bullet$ While \KU increases performance, it also increases the computing load of the mobile device. Thus, we define a dynamic control problem based on \gls{drl} (Section~\ref{sec:control}), where the controller takes as input contextual and historical information and determines (i) \KU activation, (ii) which frames are submitted for object detection and (iii) which object detection model is used. This formulation enables tight control of the delay/energy/accuracy trade-off based on contextual information. The use of \gls{drl} is motivated by (\emph{i}) the fast temporal variations of the system do not allow for long-term static optimization and require tight dynamic control; (\emph{ii}) the future statistics of the system state and performance depend on past controllable parameters (\emph{i.e., actions of the \gls{drl} agent in our framework}), so that the optimization needs to be formulated as a correlated control sequence controlling the system's state trajectory, rather than a one-shot optimization of the system parameters. We demonstrate in Section~\ref{sec:results} that performance is a function not only of ``system'' variables such as delay, but also of measurable ``content'' variables such as the dynamics of the objects, and ``algorithm'' parameters (\emph{e.g.}, the object detection model).\smallskip

$\bullet$ We refer to the resulting framework, illustrated in Fig.~\ref{fig:framework}, as \FW. We train and evaluate \FW on a real-world experimental platform. Our results demonstrate that by adapting the strategy to the context, \FW achieves superior tracking performance (4\% improvement) using considerably less power (60\% reduction in \KU activation) and channel and edge server resources (50\% reduction) compared to any non-adaptive strategy. With respect to a fixed strategy without \KU, \FW improves mAR by 20\%. Importantly, the \FW \gls{drl} agent uses significantly different control strategies for different parameters of the system (link quality) and video (target mobility), thus confirming the need for context-aware control.

\section{Related work}

Thanks to its relevance in many critical real-world applications, real-time video analytics has recently attracted significant attention. Prior work has proposed techniques to reduce the computation burden and latency of image analysis algorithms to match the resources and constraints of mobile applications, including model pruning~\cite{deepadapter}, advanced compression~\cite{yao2017deepiot} and split DNNs \cite{matsubara2019distilled}.
For the same purpose, some recent contributions apply a joint \gls{od} and \gls{ot} strategy on video streams~\cite{xu2021approxnet, approxdet}. Among others, the recent ApproxDet framework~\cite{approxdet} is one of the closest to our work. However, the latter focuses on a purely local computing scenario, where the mobile device executes both \gls{od} and \gls{ot}. In this context, ApproxDet  selects which frames are processed using \gls{od} and which using \gls{ot}, as well as some computing parameters. However, this methodology suffers from a critical issue --  frames analyzed using \gls{od} incur a large delay. As a result: (\emph{a}) the bounding boxes for those frames would become available after an excessive amount of time to support real-time applications;  (\emph{b}) during \gls{od} processing, a non-negligible number of frames would be completely disregarded; and (\emph{c}) in a real-world setting, where the captured scene evolves during \gls{od} analysis, the reference provided by \gls{od} would become obsolete, and tracking performance would significantly degrade unless slowly-changing videos were considered,  as demonstrated by our results in Section~\ref{sec:results}. Noticeably, ApproxDet only considers a subset of slowly varying videos from \textit{ILSVRC 2015 - VID} with large subjects, and only shows $95$ percentile latency.
In this paper, we propose an edge computing-based solution where \gls{od} and \gls{ot} are executed \textit{in parallel} on different machines.  Although this approach provides firm guarantees on bounding boxes delay, it makes control more challenging (\emph{e.g.}, due to the erratic behavior of the wireless channel), which we address by developing a context-aware \gls{drl} controller.  Furthermore, we introduce \KU to increase resiliency to \gls{od} delay.


In the class of \gls{od}-only solutions based on edge computing, to decrease \gls{od} latency image segmentation has been explored by recent some frameworks, including ELF~\cite{elf}. The core idea is to adapt computing based on previous \gls{od} outcomes to optimize analysis over multiple edge servers. Specifically, ELF produces a region proposal prediction, based on LSTM with attention networks, that predicts the new bounding boxes given the previous ones. Next, the frame is fragmented to distribute the load to the different edge servers based on their load. Thus, ELF focuses on remote \gls{od} only, while we propose the use of dual and parallel \gls{od}-\gls{ot}, which provides firm latency guarantees. Moreover, the scenario considered in ELF centers on load distribution across multiple edge servers over a high-capacity channel. Conversely, the key innovation of \FW is to increase resiliency to erratic and limited channel capacity, as well as to latency variations, to support \gls{ot} based on the current video content and networking contexts.

%

\glsresetall

\section{Real-Time Distributed Video Analysis}
\label{sec:problem}

In this Section, we provide an overview of the distributed video analysis scenario considered in this paper.\smallskip

\subsection{Background and System Model}
\label{sec:background}

\begin{figure}[!h]
    \centering
    \includegraphics[width=\columnwidth]{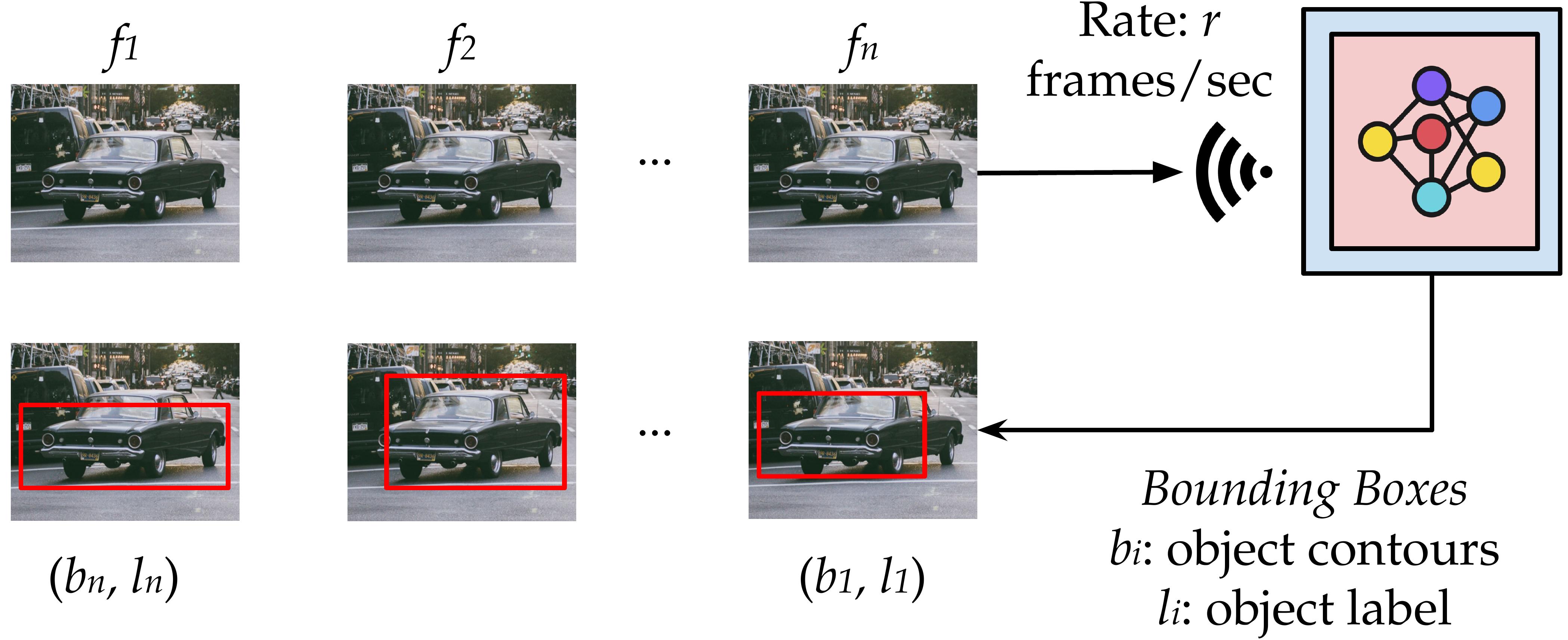}
    \caption{Object detection (OD) in edge-based systems.\vspace{-0.4cm}}
    \label{fig:edge_od}
\end{figure}

Figure \ref{fig:edge_od} depicts the edge-based object detection process under investigation. Specifically, we consider a mobile device capturing a sequence of images\footnote{In this paper, we will use the words image and frame interchangeably.} $f_1, \ldots f_{N_i}$, at a fixed rate of $r$ images per second. The general objective of the system is to analyze the images to detect objects. Specifically, each image $f_i$ is associated with a vector of object descriptors $O_i{=}(b_1,l_1,\ldots,b_{N_i},l_{N_i})_i$, where $b_j$ and $l_j$ are respectively the bounding boxes enclosing the $j$-th object in the image and its label. The bounding box is defined as the minimum rectangle enclosing all the pixels of an object, and the label is an integer corresponding to a class describing the nature of the object in a finite set. We note that the number of objects $N_i$ in the image $f_i$ is a function of the image itself.

We denote as $O_i$ the vector containing the reference ground truth. The system extracts an approximation $\hat{O}_i$ of $O_i$ using the object detection function $\phi(\cdot)$, i.e.,   $\hat{O}_i{=}\phi(f_i)$.  The quality of the approximation is defined by metrics such as \gls{map} or \gls{mar}. These metrics evaluate the quality of the bounding boxes generated by the algorithm, as well as their classification. Henceforth,  we will use \gls{mar} to measure the ability of our approach to recognize targets. This metric is based on recall, that is, the normalized number of targets correctly labeled in a single frame with an intersection over union (defined as the intersection area of the ground truth and predicted bounding boxes divided by their union area) larger than $0.5$. To compute \gls{mar}, the recall is averaged over a whole video or a portion of it.

We consider \glspl{dnn} for \gls{od}, which are the \textit{de-facto} new standard to perform object detection in real-world applications.
Many of these networks are categorized in families of networks: CenterNet Hourglass \cite{centernet}, SSD Resnet \cite{resnet}, Faster RCNN Resnet, Yolo \cite{yolov3}, just to name a few. A significant number  of these architectures are scalable, which creates a number of \glspl{dnn} that have performance proportional to their size. However, the complexity of most of these models is beyond the capabilities of current mobile devices. Even relatively powerful embedded computers such as the NVIDIA Jetson Nano we use in this paper cannot execute even medium size \gls{dnn} models for object detection due to memory constraints.  Other models are supported, but their execution requires an excessive time and significantly increases power consumption.  We report specific values in Section~\ref{sec:results} Table 1.

We then take a joint \gls{od} and \gls{ot} approach, and define two distinct functions for the estimation of $\hat{O}_i$. Formally, in addition to the \gls{od} function $\hat{O}_i{=}\sigma^{\rm od}(f_i)$, we define  the \gls{ot} function as $\hat{O}_i{=}\sigma^{\rm ot}(\hat{O}_{i-1},f_i)$.  Thus, $\sigma^{\rm ot}$ takes as input the current image as well as the estimated object descriptors associated with the previous image to leverage temporal correlation in the image stream. Different types of \gls{ot} algorithms have been proposed. Some of these methods are based on features extraction algorithms \cite{csrt_tracker} (such as Histogram of Oriented Gradients) or deep architectures (using for example siamese networks, resulting in algorithms such as GOTURN~\cite{goturn}).  Other algorithms are based on optical flow, using classical techniques such as the Lucas-Kanade point tracking. Among these, MedianFlow \cite{medianflow} takes the median of flow vectors generated to predict where the new location of the bounding box.  In this paper, we use  MedianFlow  due to its low-complexity, which satisfies the latency and resource constraints which characterize real-time applications. 

Critically, \gls{ot} algorithms rely not only on a good estimate of the object descriptors associated with the previous image, \textit{but also on a limited change in the image}. Due to the nature of these algorithms, their performance is inversely proportional to the rate with which the video changes. Expanding the neighboring region where to look for a matching set of features (extracted with DNNs or HOGs or macro-blocks of pixels) to follow fast moving objects results in higher uncertainty and consequently poorer tracking performance.  Furthermore, error accumulation and consequent target instability have been well documented \cite{CIAPARRONE202061, approxdet}. For these reasons, \gls{od} is periodically executed to ``reset'' the bounding boxes by providing a new and independent reference, which is then sequentially updated using \gls{ot} as new images are acquired~\cite{detect_and_track}.  The \gls{ot} algorithm we adopted is orders of magnitude less complex compared to \gls{od} \cite{detect_and_track}.  On the other hand, \gls{od}-designated frames still incur in large latency.
As a result, in traditional approaches such as ApproxDet \cite{approxdet}, where the mobile device executes both \gls{od} and \gls{ot}, tracking is halted while waiting for the outcome of object detection. Thus, either the incoming frames during this time are discarded, or they are buffered and processed with a larger accumulated delay. We note that in both cases the correlation between the \gls{od} reference and the images processed with \gls{ot} decreases due to the time lag.


\subsection{Edge Offloading of Object Detection}

Our core idea is to divide the video analysis into two parallel yet intermingled processes: \emph{object tracking} executed locally at the mobile device on all frames, and \emph{object detection} executed remotely at the edge server on a subset of frames $\mathcal{E}{\subseteq}\{0,1,\ldots\}$. The key advantages of this strategy are the following: (i) the edge server has a larger computing power compared to the mobile device, so that the execution time of object detection is reduced, (ii) the two processes can be fully executed in parallel without sharing resources, and (iii) the overall energy consumption at the mobile device is reduced.
However, offloading the execution of object detection to the edge server requires the transportation of the image to be analyzed over a wireless link. In many real-world settings, the channel capacity is constrained and erratic (\emph{e.g.}, autonomous vehicles, millimeter wave communications, \emph{etc.}). Moreover, offloading may result in channel congestion, thus  increasing delay and amplifying data rate instability. Thus, it becomes necessary to parsimoniously send frames to the edge server.

Let us denote with $\Delta^{\rm od}_i$, $i{\in}\mathcal{E}$, the total time from the capture of the image $i$ to the reception of the vector $\hat{O}^{\rm od}_i$ when the frame is sent to the edge server. $\Delta^{od}_i$ is the sum of communication time and computing time. The former is a function of the perceived data rate and the number of bytes used to represent the image. The latter is a function of the \gls{od} \gls{dnn} model used at the edge server and its computing power. Both delay components are time-varying as they depend on channel and system parameters. We denote as $\Delta^{\rm ot}_i$ the time from the generation of the frame $i$ to the availability of the vector $\hat{O}^{\rm ot}_i$. Due to the low complexity of the \gls{ot} algorithm we adopt, we assume that the time $\Delta^{\rm ot}_i$ is fixed and smaller than the inter-frame generation $1/r$.
We note that as object tracking is applied to all the images, an estimate of the bounding boxes for all the frames is readily available to the mobile device.

Consider a frame $i \in \mathcal{E}$, which is both processed locally using \gls{ot} and sent to the edge server for \gls{od}. In a short amount of time, $\hat{O}^{\rm ot}_i$ becomes available as \gls{ot} is executed using one of the available vectors $\hat{O}_{i-1}$. 
Then, $\hat{O}^{\rm ot}_i$ can be used as reference for the successive frame and so on. When $\hat{O}^{\rm od}_i$ from \gls{od} is received, the bounding boxes and labels are used as reference for the \gls{ot} function $\sigma^{\rm ot}(\cdot)$ applied to frame $i{+}\lceil \Delta^{\rm od}_i \times r \rceil$.
Since the \gls{ot} reference is outdated,  the tracking performance may degrade. While it is possible to continue using the reference obtained from object tracking on the previous frame, due to error accumulation in object tracking, a periodic \emph{refresh} is needed.
In Section~\ref{sec:results}, we characterize this degradation as a function of key video and system parameters.


\section{The \FW Framework}
\label{sec:framework}

\begin{figure}[h]
    \centering
    \includegraphics[width=0.75\columnwidth]{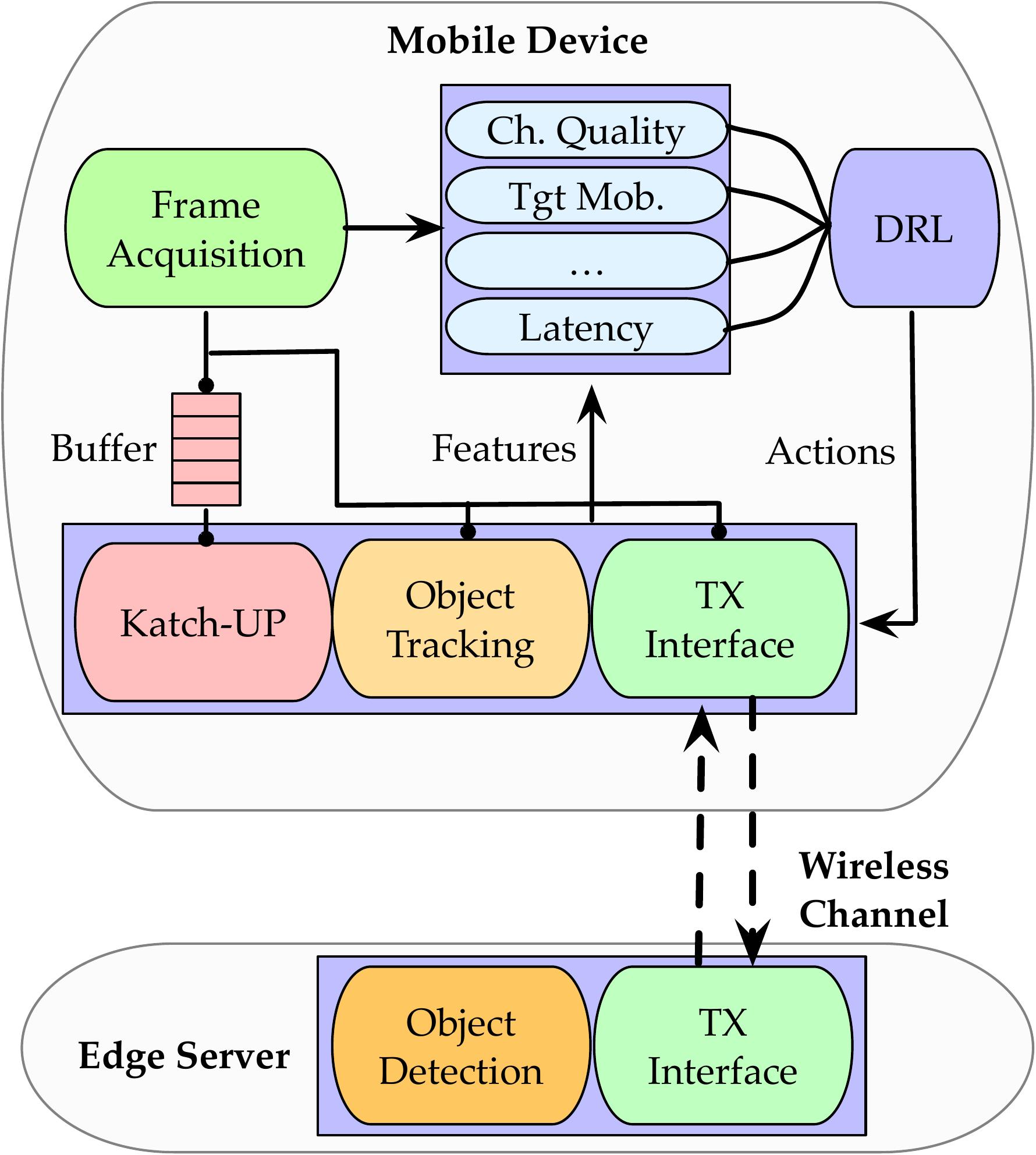}
    \caption{Main components of \FW. At the mobile device, the DRL module (state extraction and controller) determines which frames are sent for \gls{od} and what model is used for their analysis, and the activation of \KU. To support these actions, frames are internally routed to the different modules (object tracker, \KU and \KU buffer and TX interface) to support these functions. The edge server performs object detection on the received frames using the model indicated by the \gls{drl} controller. \vspace{-0.6cm}}
    \label{fig:smartdet}
\end{figure}

Let us now present the two key contributions of this paper: (\emph{i}) \KU: a methodology to make the distributed video analysis system less sensitive to object detection delay, and (\emph{ii}) \FW, a real-time control engine to optimize the tradeoff between performance and resource usage. The schematics of \FW are depicted in Fig.~\ref{fig:smartdet}. The core of \FW is a \gls{drl} agent that controls which images are sent to the edge server for \gls{od} and which model is used to analyze them, as well as whether \KU is used or not. To support these decisions, images are internally routed to the main modules: object tracker, \KU (and \KU buffer) and the transmission interface. The edge server receives the frames, and analyze them using the \gls{od} model indicated by the \gls{drl} controller, then returning the estimated bounding boxes and labels to the mobile device. A critical module of \FW is the state extractor, that builds state features from variables, parameters and data received by the other modules.

\subsection{The \KU Smart Tracking Algorithm}
\label{sec:KU}

\begin{figure}[h]
    \centering
    \includegraphics[width=\columnwidth]{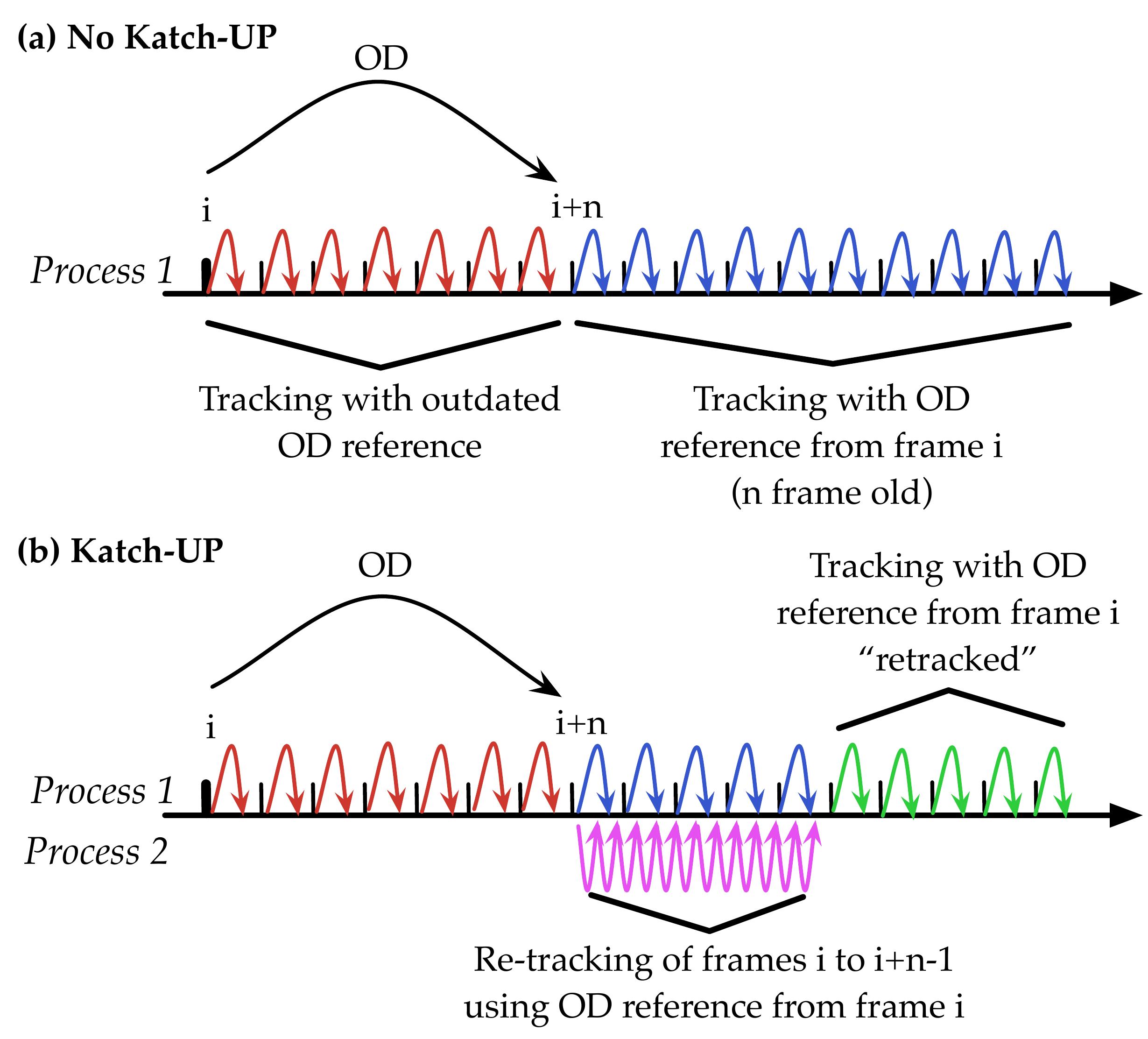}
    \caption{Tracking process with and without \KU referring to the explanation in Section~\ref{sec:KU}.\vspace{-0.4cm}}
    \label{fig:KU}
\end{figure}

As mentioned earlier, one of the main issues of \gls{od} edge offloading is the feeding of outdated references to the \gls{ot} due to the communication and computing delay. As a result, applying \gls{ot} algorithms to frames that might substantially differ will entail large errors due to the high uncertainty of the transposing vectors \cite{medianflow}. To mitigate the effect described above, we propose \KU. Our intuition is simple yet effective: when a vector $\hat{O}^{\rm od}_i$ from \gls{od} applied to frame $i$ is received, the mobile device re-executes the tracker on the frames starting $i+1$ until the process ``catches up'' with the primary tracking process. To make an example (represented in Fig.~\ref{fig:KU}), assume frame $i$ is sent to the edge server and the corresponding reference $\hat{O}^{\rm od}_i$ is received right before frame $i+n$ is acquired. During this time, \gls{ot} is applied to frames from $i$ to $i+n-1$ based on the reference available at the time (\emph{i.e.}, the outcome of tracking applied to the previous frame based on a chain of tracking started from an older object detection reference). In \KU, when $\hat{O}^{\rm od}_i$ is received the tracking process is duplicated. \emph{Process 1} continues to analyze incoming frames $i+n,i+n+1,i+n+2,\ldots$ using the reference vector $\hat{O}^{\rm ot}_{j-1}$ to perform tracking on frame $j$. \emph{Process 2} restarts the tracking of frames $i+1,i+2,\ldots$ taking $\hat{O}^{\rm od}_i$ as a starting point to build the sequence $\hat{O}^{\rm ot}_{i+1},\hat{O}^{\rm ot}_{i+2}, \ldots$. Process 2 is executed as the maximum possible speed, meaning that tracking is continuous, rather than based on the frame arrival timing. Thus, Process 2 proceeds faster than Process 1, and eventually \emph{catches up} with the latter one. Meaning, Process 2 and Process 1 generate a bounding box vector with the same index. At that point, Process 2 is terminated, and Process 1 continues using as reference the latest outcome of Process 2.

The key advantage of \KU is that the sequence $\hat{O}^{\rm ot}_{i+1},\hat{O}^{\rm ot}_{i+2}, \ldots$ generated by Process 2 is more accurate compared to that generated by Process 1, as the former is based on a more recent reference from object detection.
Thus, \KU boosts tracking performance (as demonstrated in Section~\ref{sec:results}), but also increases the computing load at the mobile device, as well as memory usage as some already processed frames need to be buffered.

\subsection{Real-Time DRL-Based Control in \FW}
\label{sec:control}

\textbf{Motivation.}~The system performance in terms of latency/accuracy/energy is determined by several factors and parameters, including: (\emph{i}) whether or not to activate \KU, (\emph{ii}) how many and which frames are to be sent to the edge server for object detection, and (\emph{iii}) which object detection model to use. For example, if \KU is active,  the quality of tracking improves, but energy consumption at the mobile device increases. If more frames are sent to the edge server, then tracking has more frequent references, but channel load -- and thus possibly communication latency -- increases as well as server load -- and thus possibly computing latency. If a more complex model is used, then the reference quality for tracking improves, but so does the latency to receive the bounding boxes. Intuitively, the optimal point is determined by several variables. For instance, if the channel has a large capacity, then transmitting more frames will not significantly affect overall delay, while possibly improving tracking. Moreover, if the communication delay is small, then the use of a more complex model, with a larger execution time, may be advantageous. Conversely, if the channel capacity is small, then using a less complex model may result in a tolerable overall latency. Notice that different object detection models take as input images of different size, and may be more or less sensitive to compression. In other words, there might be a dependency between $\Delta_i^{\rm od}$ and the model used and desired accuracy. Importantly, these trade-offs are greatly influenced by the parameters of the video itself. Moreover, the mobility of the targets in the video influences the optimal parameter choice, where fast changes may require low-latency, more frequent, object detection These tradeoffs are detailed in Section~\ref{sec:results}. The decisions in (\emph{i})--(\emph{iii}) will become the knobs used by \FW to control the tradeoff between performance and resource usage.

{\bf Why \gls{drl}?} We formulate our decision making process as a dynamic control problem, where a controller selects real-time actions at a fine temporal granularity based on the perceived state of the system and context. This approach is motivated by the time-varying nature of the system we consider, as well as by the correlation between current decisions and future  states of the system. For instance, the period determines the sampling instants of the state, as well as how outdated the reference from \gls{od} is when applying \gls{ot} to future frames.
Thus, a reinforcement learning approach is the most suitable to solve our problem.  Moreover, we observe that the state space is extensive, and the features are heterogeneous in nature, where some of them are continuous. A traditional ``tabular'' Q-Learning approach in practice would require the quantization of all features to generate a discrete space. The resulting state space would be either too large to handle using direct recursive estimation, or poorly representative. Therefore, a Deep Q-Learning approach is the natural choice for our problem.\smallskip

\textbf{DRL Algorithm.}~In \gls{drl}, the controller selects an action $u{\in}\mathcal{U}$ based on the current state $s{\in}\mathcal{S}$, where $\mathcal{U}$ and $\mathcal{S}$ are the action and state space, respectively. We synchronize the state update and action selection with the generation of images that are sent to the edge server for object detection. We index these instants with $t{=}1,2,\ldots$, and denote the state and action at time $t$ as $s_t$ and $u_t$, respectively.
We adopt a Q-Learning formulation and define the function $\omega(\cdot)$ as $Q(u_{t+1},s_{t+1}){=}\omega(s_t)$, where
\begin{align}
    Q(u_{t+1},s_{t+1}) &= \mathit{E}_{s_{t+1}|s_t,u_t}\left[\mathit{E}_{r_{t+1}|s_{t+1},u_t,s_t}\left[ r_{t+1} | s_{t+1},u_t,s_t| \right]\right]\nonumber \\ & + \gamma \max_{u^{\prime}} \mathit{E}_{s_{t+1}|s_t,u_t}\left[ Q(s_{t+1},u^{\prime}) \right],
    \label{eq:QV}
\end{align}
where $r_t$ is the reward accrued at time $t$. Thus, the Q-value $Q(s,u)$ captures the long-term -- discounted -- reward associated with taking action $u$ in state $s$. We refer the reader to \cite{sutton2018, henderson2018deep} for a comprehensive discussion on \gls{drl}.

In the following, we define the action space, state space, cost function and network architecture of the DRL agent.
We remark that the time granularity of state update and decision making is not the same as that of frame generation, as decisions are made only when a frame is sent for object detection. This also reduces the computation burden to the mobile device.

\subsubsection{\textbf{Action Space}}

We define the action $u(t)$ as the vector $(k(t),p(t),m(t))$, where (\emph{a}) $k(t){\in}\{0,1\}$ determines whether \KU will be used when the outcome $\hat{O}^{\rm od}_{i_t}$ is received;
(\emph{b}) $p(t)\in\{1,\ldots, P\}$ is a variable controlling the object detection period, that is, the number of frames until the next frame is sent to the edge server for object detection; (\emph{c}) the variable $m(t){\in}\{0,\ldots,M\}$ determines which model is used for object detection of the next frame sent.

\subsubsection{\textbf{Reward Function}}

We define the reward as function of the state and action, and not of the temporal index $t$. Rewards refer to sequences of frames in between \gls{od}. Next, we define the reward function $R(s,u)$ as the composition of the following metrics: 
(\emph{i}) mAR ($R_1(s,u)$): counts the percentage of the targets we correctly track with respect to the total number of targets; (\emph{ii}) \emph{KU usage} ($R_2(s,u)$): is the fraction of frames processed using \KU, (\emph{iii}) \emph{Period} ($R_3(s,u)$): is the number of frames until the next scheduled \gls{od} (normalized to the maximum period).
An exhaustive description is not provided due to space constraints.
Note that we do not include \gls{od} latency directly in the reward function as it is not a direct application metric.
Indeed, the latency perceived by the application is that of \gls{ot}.
However, \gls{od} latency influences tracking performance, and thus \gls{mar}.
We then define $R(s,u)$ as the weighted sum
\begin{equation}
\label{eq:reward}
    R(s,u)= \alpha_1 R_1(s,u) + \alpha_2 R_2(s,u) + \alpha_3 R_3(s,u).
\end{equation}

\subsubsection{\textbf{State Space}}

The state $s{\in}\mathcal{S}$ is designed to provide information to the controller to make decisions on the action to be selected. At a high level, to estimate the $Q$ function, the controller needs to predict some characteristics of the video and surrounding system, and connect them to a reward given the action. To this purpose, the \gls{dnn} $\omega$ embedded in the controller implicitly builds a model for the temporal evolution of parameters such as video characteristics and channel. Next, we include in the state a set of \emph{features} over a window of $N$ past decision instants.
That is, at a decision instant $t$, we include these features computed at $t{-}N{+}1, t{-}N{+}2,\ldots,t$.
We can group these features as follows:\\
\noindent
(\emph{a}) \emph{Contextual}: this includes image size and the center-to-center distance between targets as a proxy on how quickly objects are moving in the frame;\\
\noindent
(\emph{b}) \emph{Self-awareness}: the latency incurred by the frames sent for object detection, KU usage (same as in the reward computation), and the selected action vector;\\
\noindent
(\emph{c}) \emph{Self-evaluation}: \gls{iou} between the running tracking and the received detection.

Examples of ``\emph{correlations}'' that the agent will need to learn by experience include the temporal correlation of video change rate and \gls{od} latency. Note that the latter depends not only on the time varying channel capacity and server load, but also on the model used for \gls{od}, so that the agent \textit{needs to implicitly learn model-to-model latency maps.} The agent also needs to learn the \textit{non-trivial relationship between \gls{od}, \KU, the (future) latency and video parameters and \gls{mar}}.

\begin{figure}[h]
    \centering
    \includegraphics[width=.85\columnwidth]{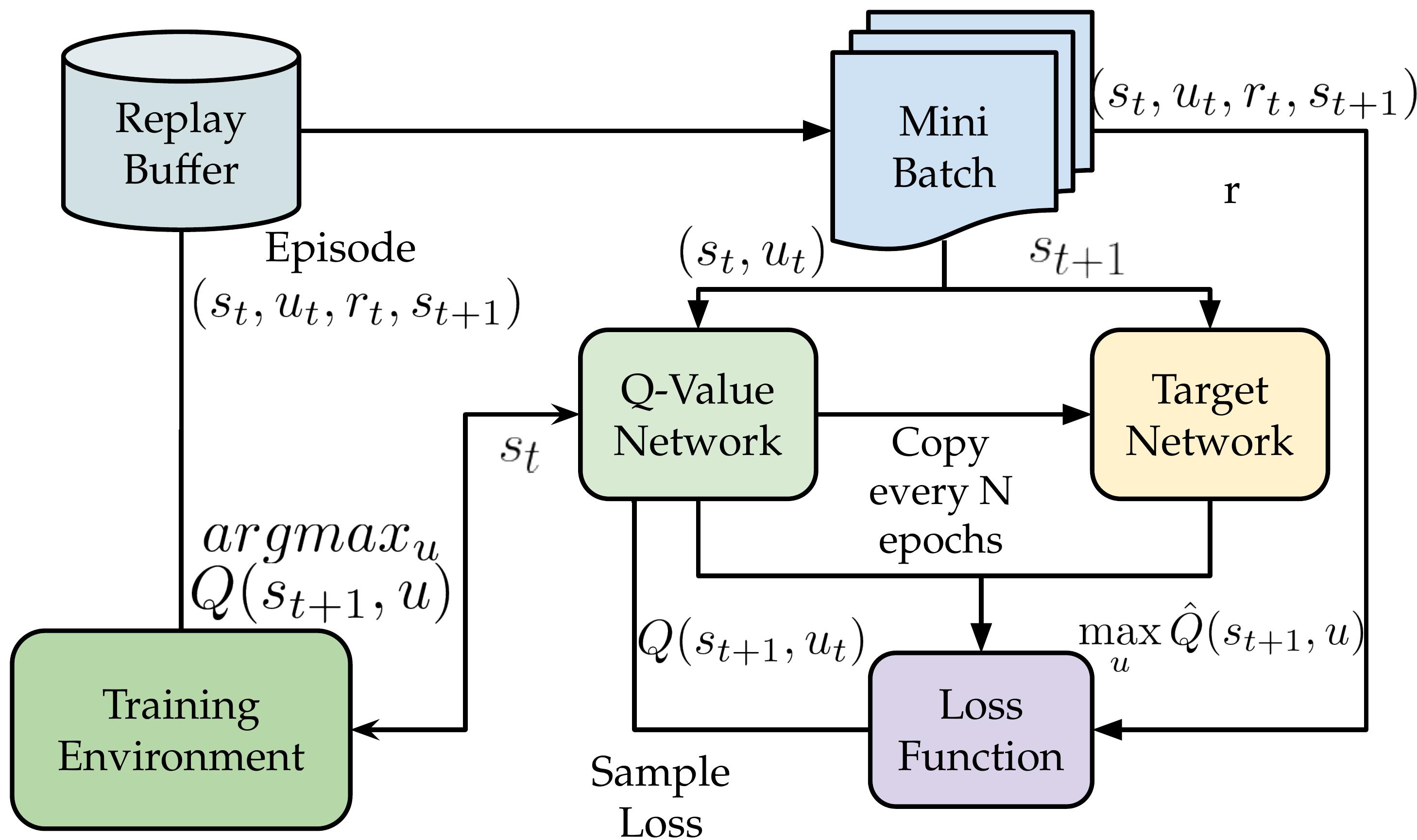}
    \caption{Deep Q Neural network with replay buffer architecture.}
    \label{fig:DDQL}
\end{figure}
\subsubsection{Implementation details}
We use a network composed of $5$ layers of $64$, fully connected ReLu activated nodes. The low-complexity of the network easily fits the constraints of mobile devices. In the platform we consider, the network can be executed at up to $10Hz$ increasing the power consumption by only $<1\%$. We adopt a double Deep-Q Learning structure (see Fig.~\ref{fig:DDQL}) to train the network \cite{ddqn}, where the Q-value network learns the relationship between the input state and the output Q-values (one for each action) by using the target network for the $Q(s_{t+1}, u)$ value, and  Eq.~\ref{eq:QV} to combine it with the reward. To balance exploration-exploitation, we use the, effective and stable, $\epsilon$-greedy scheme~\cite{sutton2018}.

\section{Experimental Results}
\label{sec:results}

We describe our experimental setup and dataset generation process in Section \ref{sec:dataset}. Then, we discuss our experimental trade-off analysis in Section \ref{sec:results1}. Finally, we evaluate and compare \FW against baselines in Section \ref{sec:smartdet_results}.

\subsection{Experimental Setup and Data Collection}\label{sec:dataset}

All the experiments were performed indoor in a campus setting. As mobile device, we use an NVIDIA Jetson Nano, quad-core ARM $1.9$GHz CPU and mounting a $128$-core GPU operating at $0.95$GHz, with performance comparable to current generation mobile phones and small autonomous vehicles \cite{skydio}. As edge server, we use a ThinkPad P72 with hexa-core CPU operating up to $4.3$GHz, 32GB of memory and NVIDIA GPU Quadro P600 that has $384$ cores operating at $1.45$GHz, and a custom server, mounting 6 core CPU running up to $4.00$ GHz, 32 GB of RAM, GPU GTX 980 Ti with 2816 cores at $1.4$GHz. We set up the laptop in hotspot mode, using its Wireless-AC 9560 card, to which the mobile device connects using Realtek WiFi dongle supporting IEEE 802.11n.

\begin{figure}[h]
    \centering
    \includegraphics[width=0.7\columnwidth]{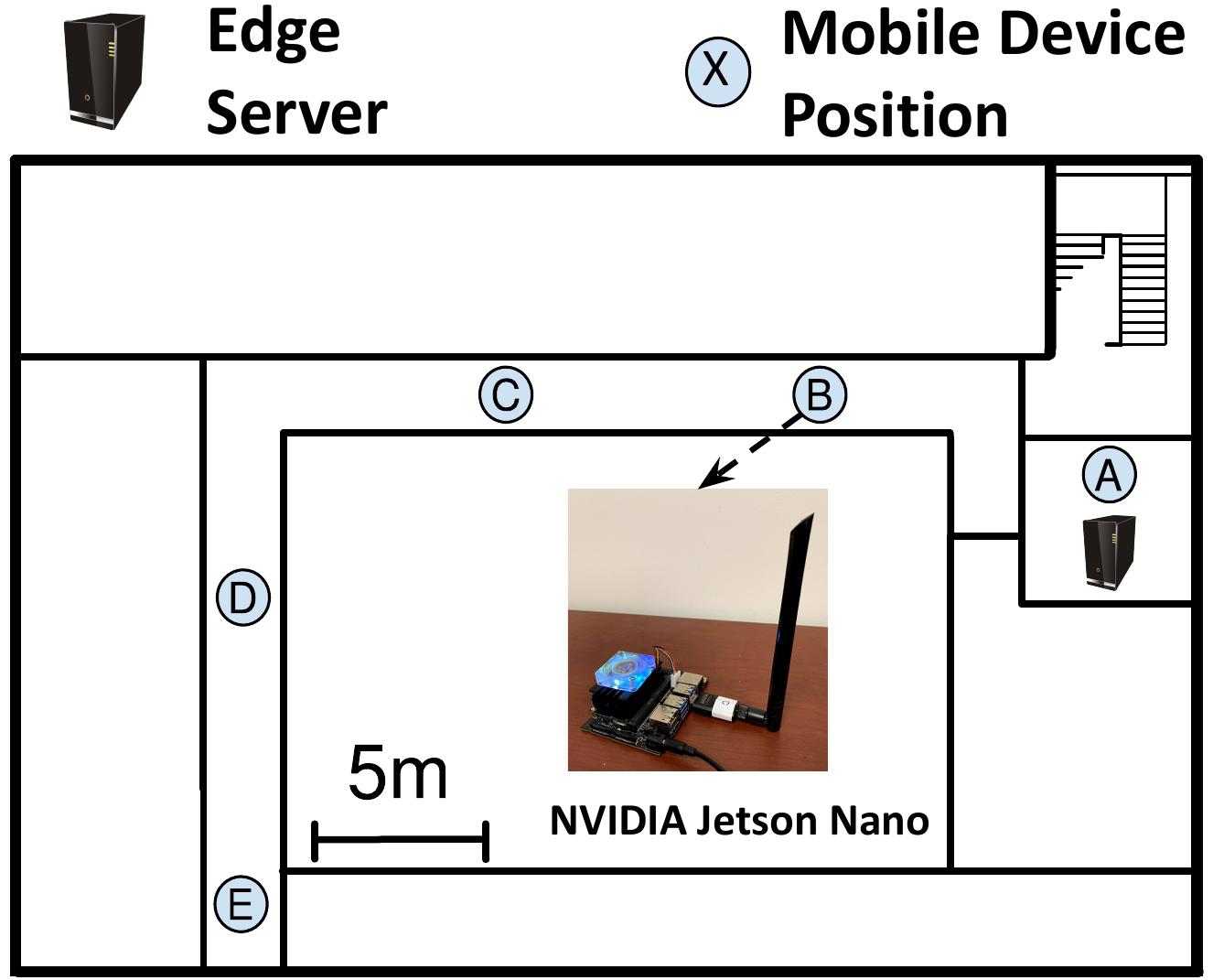}
    \caption{Representation of the experimental environment. Mobile device: NVidia Jetson Nano; Server: ThinkPad P72 and Server with GTX 980Ti GPU.}
    \label{fig:exp_setup}
\end{figure}

\begin{figure*}[t]
\includegraphics[width=\textwidth]{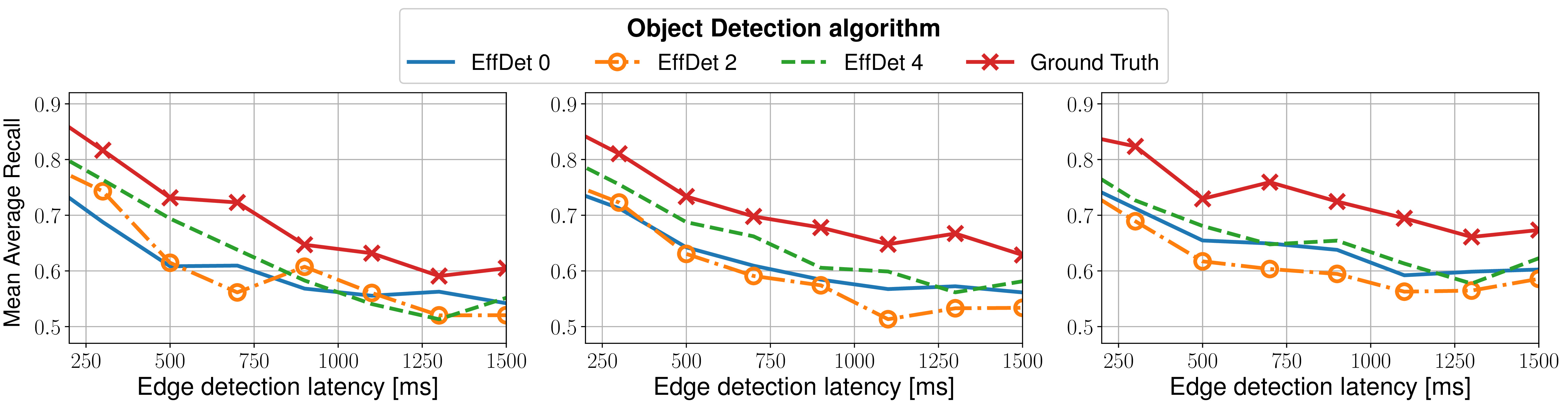} 
\caption{mAR as a function of \gls{od} latency for different periods ($0.5$, $1$, $1.5$s) for different \gls{od} models. \KU OFF.}
\label{fig:prelim1}
\end{figure*}

\begin{figure*}[t]
\includegraphics[width=\textwidth]{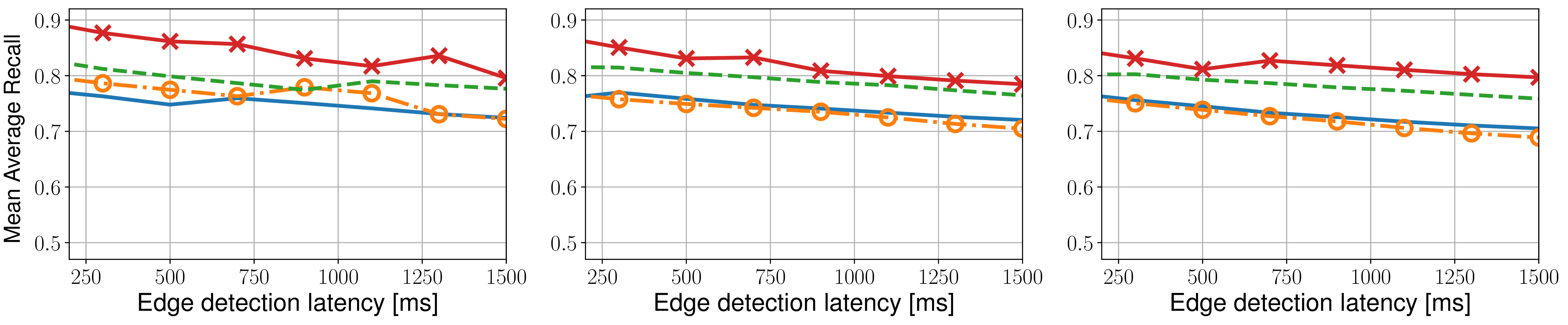}
\caption{mAR as a function of \gls{od} latency for different periods ($0.5$, $1$, $1.5$s) for different \gls{od} models. \KU ON.}
\label{fig:prelim2}
\end{figure*}

We perform our evaluation on ILSVRC2015-VID dataset, which we send over the network in order to collect the network latency in different link quality conditions, and resizing images corresponding to the optimal input of each object detection. Fig.~\ref{fig:exp_setup} shows the topologies used in the experiments.
We load the EfficientDet 0, 2, 4 models at the server, and associate them with control actions. 
We adopt MedianFlow as the object tracking algorithm to allow the mobile device to execute tracking in real time. In our preliminary evaluation, the DNN-based tracker GOTURN and CSRT (a common alternative) would achieve a frame rate below $2$fps and $3$fps, respectively.

Table~\ref{table:prelim1} reports the execution time of the various EfficientDet models on the laptop and server, and their input image size. We can see a progressive increase of execution time as the model's complexity increases. On the Laptop, EfficientDet 0 takes $0.12$s, whereas EfficientDet 4 takes almost $10$ times as much ($1.08$s). In the server, besides the lower execution times, we observe a less steep progression, where the execution of EfficientDet 4 takes about $4$ times longer than that of EfficientDet 0.

\begin{table}
\centering
\begin{tabular}{c|c|c||c}
 \hline
 Model/Server type & Laptop [s] & Server [s] & Avg. Image Size [kB] \\ 
 \hline
 D0 & 0.12  & 0.089 & 52.15 \\ 
 D1 & 0.215 & 0.11 & 69.8 \\ 
 D2 & 0.33  & 0.16  & 93.3 \\ 
 D3 & 0.59  & 0.255 & 116.3 \\ 
 D4 & 1.08  & 0.4   & 138.8 \\ 
 \hline
\end{tabular}
\caption{Execution time and image size for the various EfficientDet models. Processing units available at Laptop: Quadro P600; and at Server: GTX 980 Ti.}
\label{table:prelim1}
\vspace{-0.2cm}
\end{table}

We evaluated the instantaneous power consumption (on the Jetson Nano) of the \gls{ot} algorithm when the \KU is off and on. In the former case, the power consumption is $3512\pm242 mW$, whereas in the latter increases to $3939\pm459 mW$. 
Thus, the \gls{mar} improvement granted by the \KU technique comes at the price of an increase of power consumption of about $11$\%. 
It is then critical to activate the \KU only when necessary.
We standardize the three components of the reward function so that all changes in the actions' outcome across the reward metrics are similarly reflected in the feedback signal (as defined in Eq.~\ref{eq:reward}). 

\subsection{Tradeoff and Trends Analysis}
\label{sec:results1}

We first analyze the major trends in the system. The objective is to illustrate the key tradeoffs that will drive the controller actions.
Fig.~\ref{fig:prelim1} shows the \gls{mar} as a function of the \gls{od} latency when the \KU is inactive. The different lines correspond to various EfficientDet models ($0$, $2$ and $4$) and the ground truth. Here, the latency is abstracted from channel and system parameters. The different plots correspond to different \gls{od} periods ($0.5$, $1$ and $1.5$ seconds). The degradation of \gls{mar} as the latency increases is apparent: in the system without \KU, even when the ground truth is available for a large number of frames ($1$ every $5$), the \gls{mar} rapidly goes from $8.5$ to $6.5$ ($23$\% decrease) as the latency goes from $250$ms to $1500$ms. A similar decrease is observed when EfficientDet models are used, and for different periods. Note that executing EfficientDet 2, a medium complexity model in the considered set, takes $0.33$ and $0.16$s on the laptop and server, respectively, so that a very small latency is not expected even in ideal channel conditions.   We notice that in general a larger period -- that is, fewer frames are sent to the edge server for object detection -- results in a less pronounced, performance degradation.

Fig.~\ref{fig:prelim2} shows the same trends when the \KU is ON. Notably, besides increasing \gls{mar} in general, the \KU makes the \gls{mar} much less sensitive to latency. When the latency is above $0.5$ seconds, the superior performance of \KU is manifest. For a latency of $1500$ms and period $0.5$, the \gls{mar} goes from $0.5$ to $0.7$ when EfficientDet 0 or 2 is used, from $0.5$ to $0.77$ when EfficientDet 4 is used, and from $0.6$ to $0.8$ when using the ground truth.  Thus, as the latency of object detection increases, the controller can resort to \KU to maintain a high \gls{mar}. We note that while the difference between the various models is minimal when the \KU is off, the activation of \KU makes the difference between EfficientDet 0/2 and 4 perceivable, and the controller may leverage this difference when selecting the video analysis configuration.




\begin{figure}[t]
     \begin{subfigure}[b]{0.48\columnwidth}
         \centering
         \includegraphics[width=\textwidth,trim={5mm 5mm 5mm 5mm},clip]{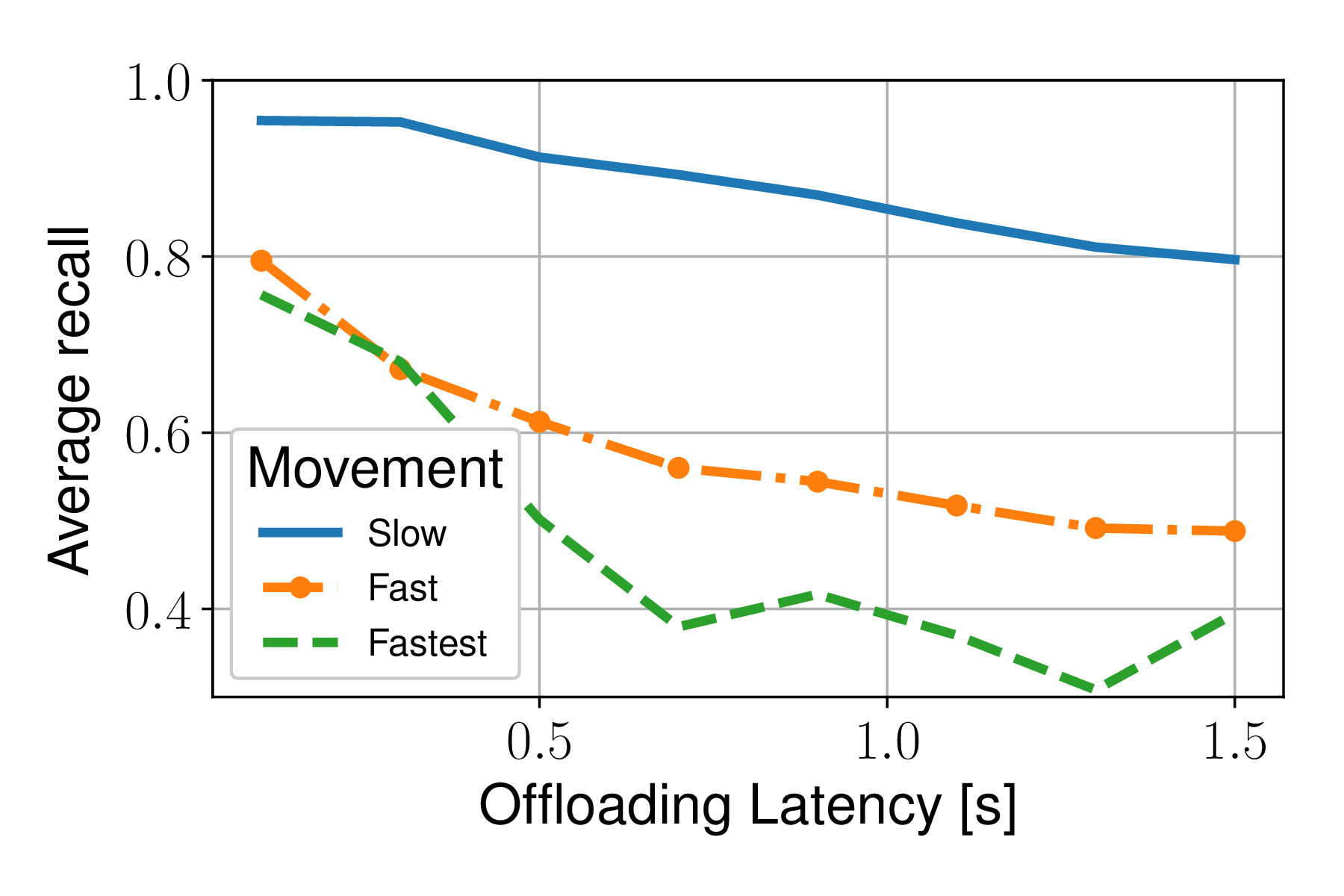}
         \caption{Katch-Up OFF}
     \end{subfigure}
     \begin{subfigure}[b]{0.48\columnwidth}
         \centering
         \includegraphics[width=\textwidth,trim={5mm 5mm 5mm 5mm},clip]{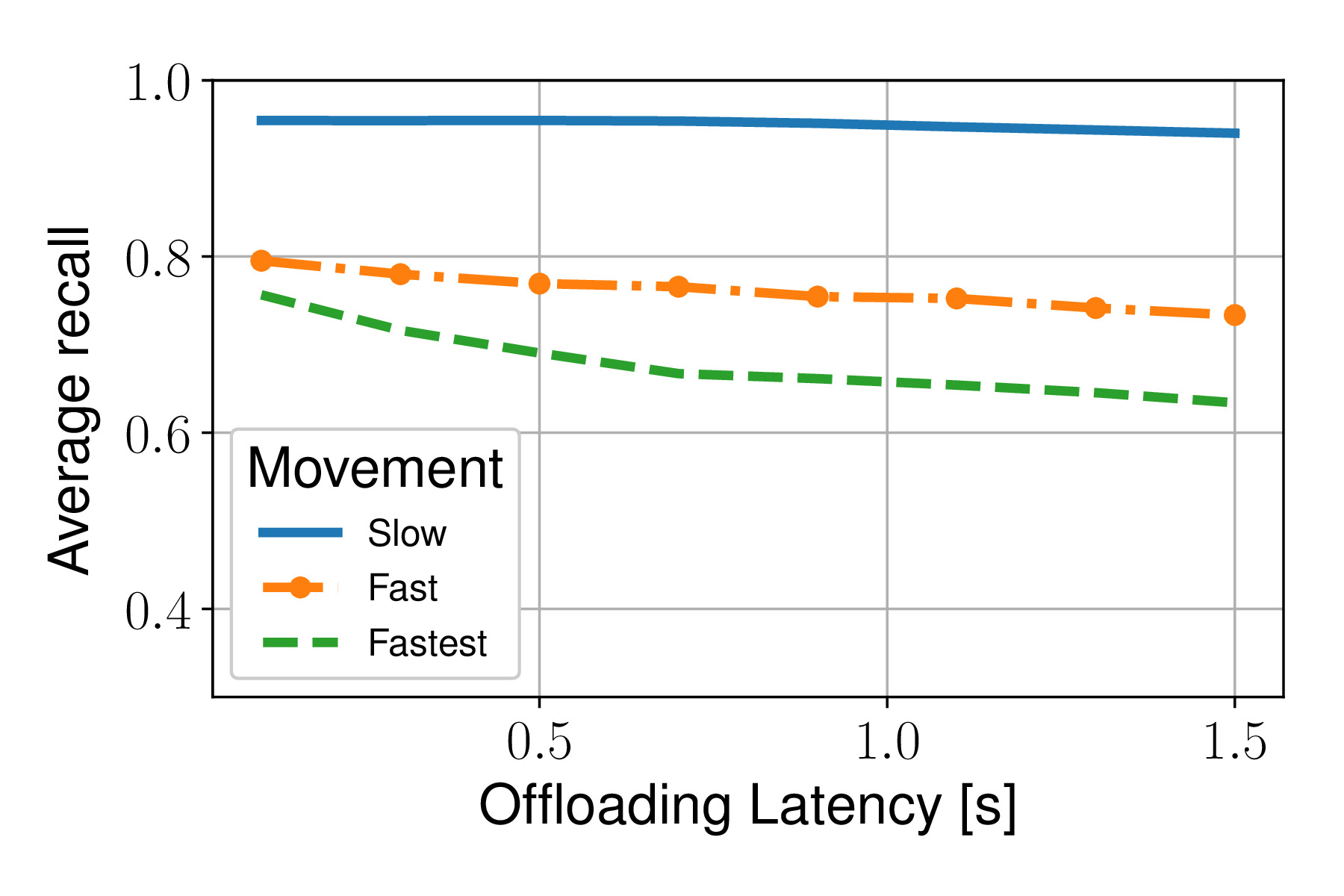}
         \caption{Katch-Up ON}
     \end{subfigure}
        \caption{mAR as a function of object detection latency for different video motion speeds. Period: $1500ms$, EffDet: 2, \gls{mar} is normalized to object detection only performance.\vspace{-4mm}}
        \label{fig:prelim3}
\end{figure}

We now analyze the effect of video parameters on \gls{mar}. Fig.~\ref{fig:prelim3} shows the \gls{mar} as a function of the \gls{od} latency for different video change rate, which we measure as the change in position of the targets, when the \KU is off (a) and on (b). 
In the figure, EfficientDet 2 is used, and the period is set to $1500ms$. Here, we classify videos as \emph{slow}, \emph{fast}, \emph{fastest} corresponding to change of position of $<5\%$ $<20\%$ or more than $20\%$ of the frame width.
The impact of motion parameters on \gls{mar} is apparent. For \KU off, at $0.5$s latency the three classes have \gls{mar} of $0.92$, $0.6$ and $0.5$, respectively. The difference increases with the latency. We can observe how \KU increases the performance and makes it less sensitive to latency for all the classes. These results demonstrate how control needs to take into account video parameters to balance energy expense and performance.

\begin{figure}[t]
     \begin{subfigure}[b]{0.48\columnwidth}
         \centering
         \includegraphics[width=\textwidth]{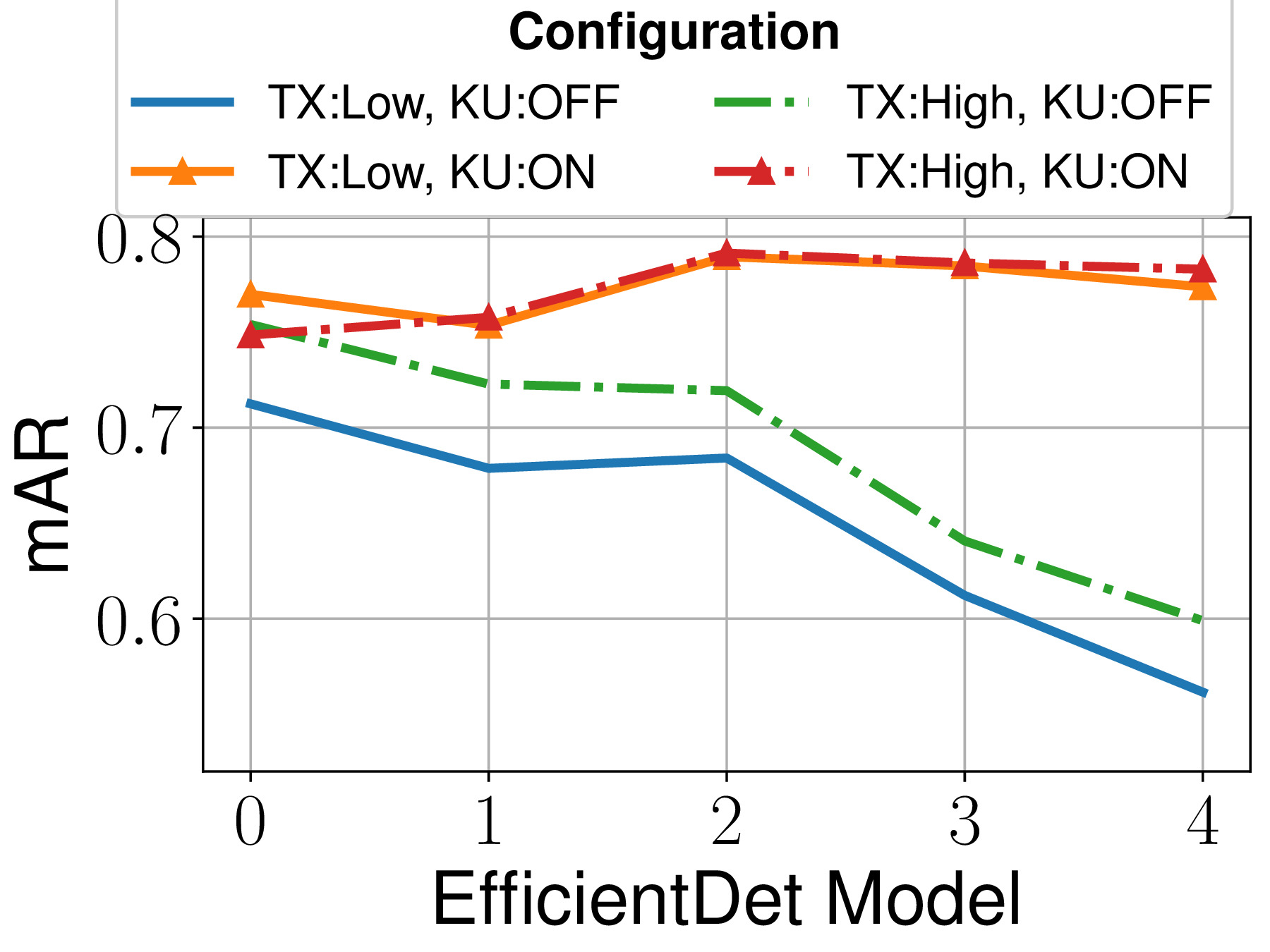}
         \caption{Laptop}
     \end{subfigure}
     \begin{subfigure}[b]{0.48\columnwidth}
         \centering
         \includegraphics[width=\textwidth]{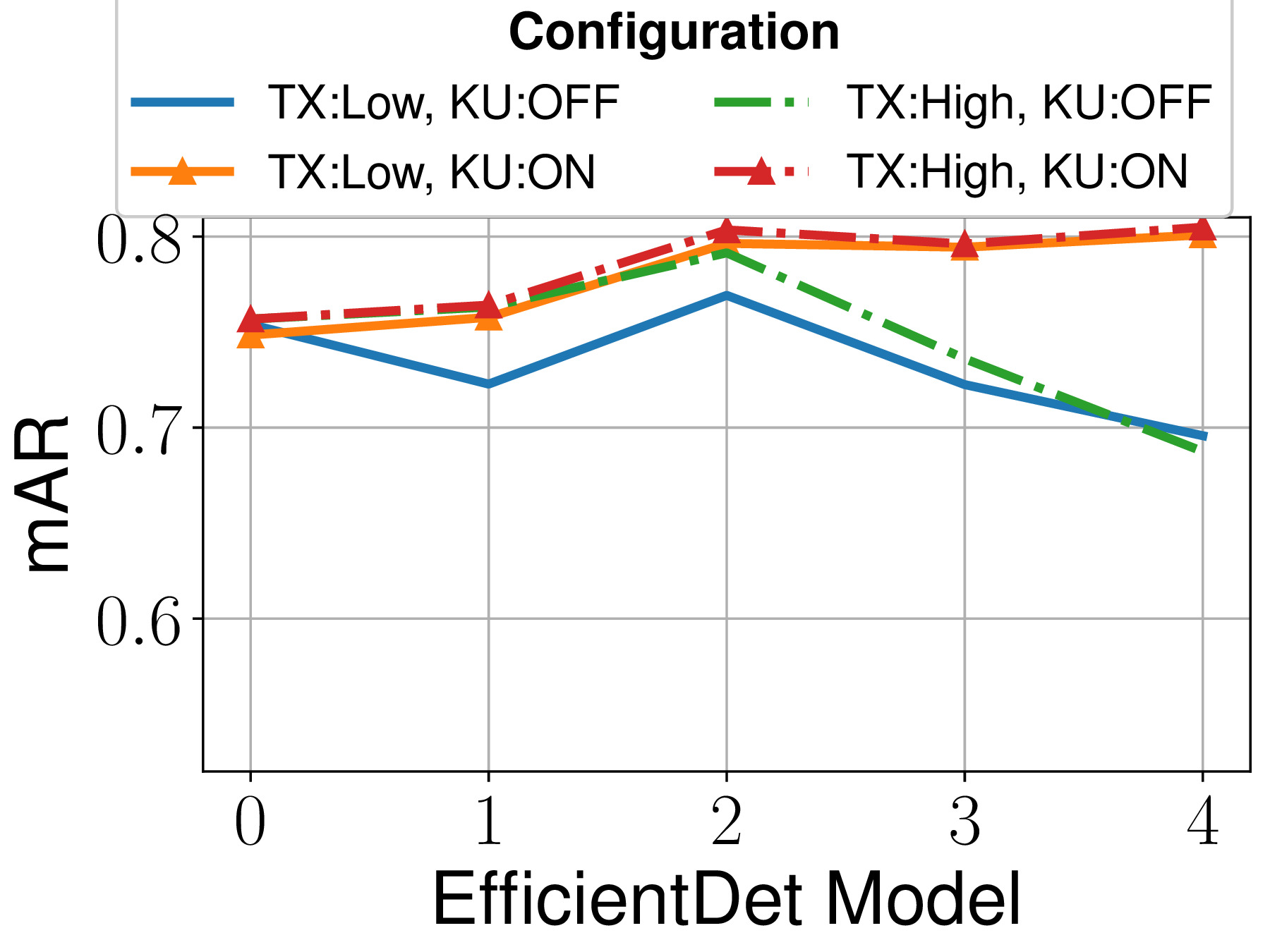}
         \caption{Server}
     \end{subfigure}
        \caption{mAR as a function of the model for different WiFi quality (Low=20, High=44), Katch-Up (ON, OFF), and Server (Laptop and Server).\vspace{-4mm}}
        \label{fig:prelim4}
\end{figure}

\begin{figure*}
     \begin{subfigure}[t]{0.33\textwidth}
     \centering
         \includegraphics[width=\textwidth,trim={5mm 5mm 5mm 5mm},clip]{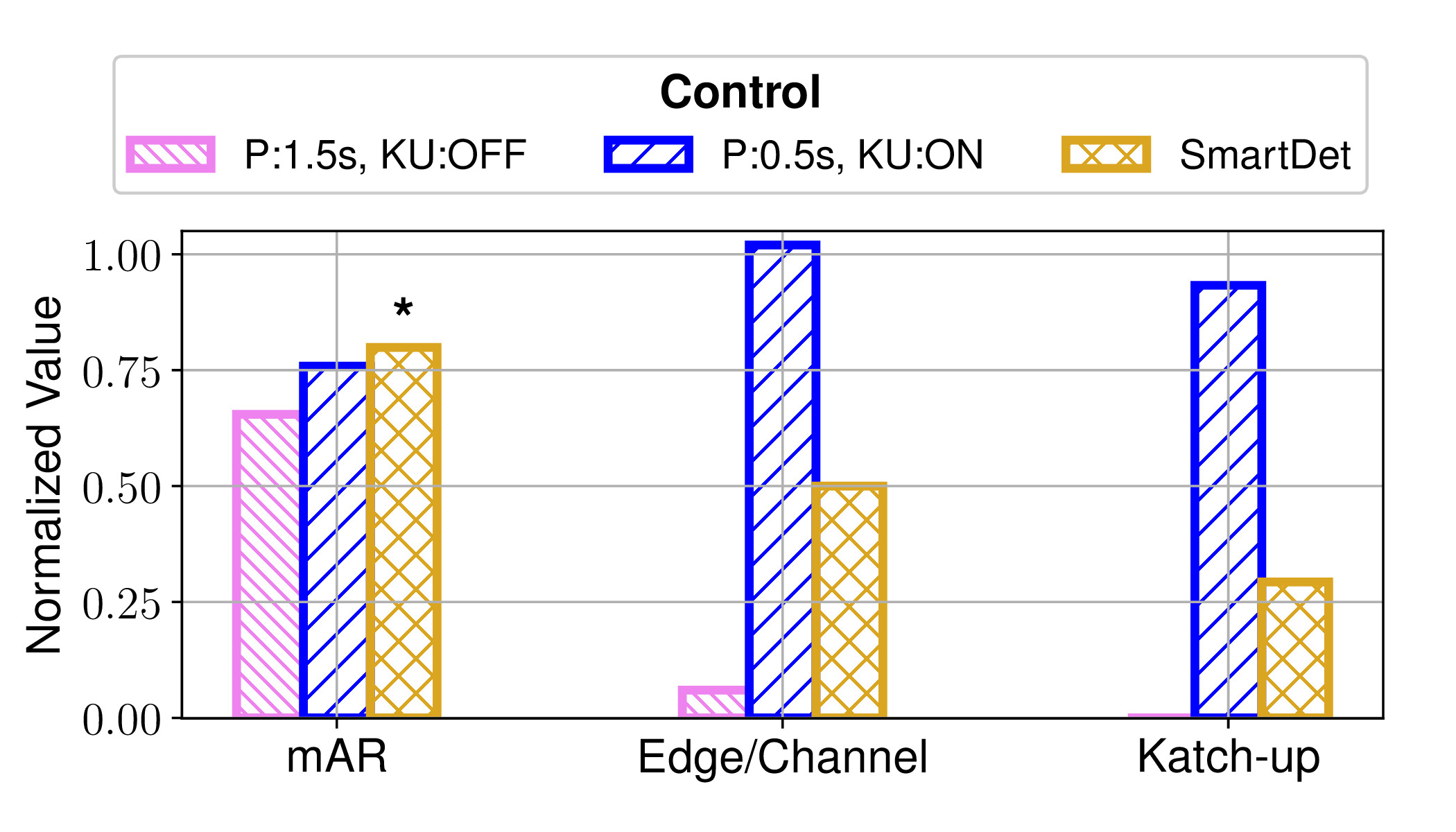}
 \caption{}
 \label{fig:comparison}
     \end{subfigure}
     \begin{subfigure}[t]{0.33\textwidth}
         \centering
         \includegraphics[width=\textwidth,trim={5mm 5mm 5mm 5mm},clip]{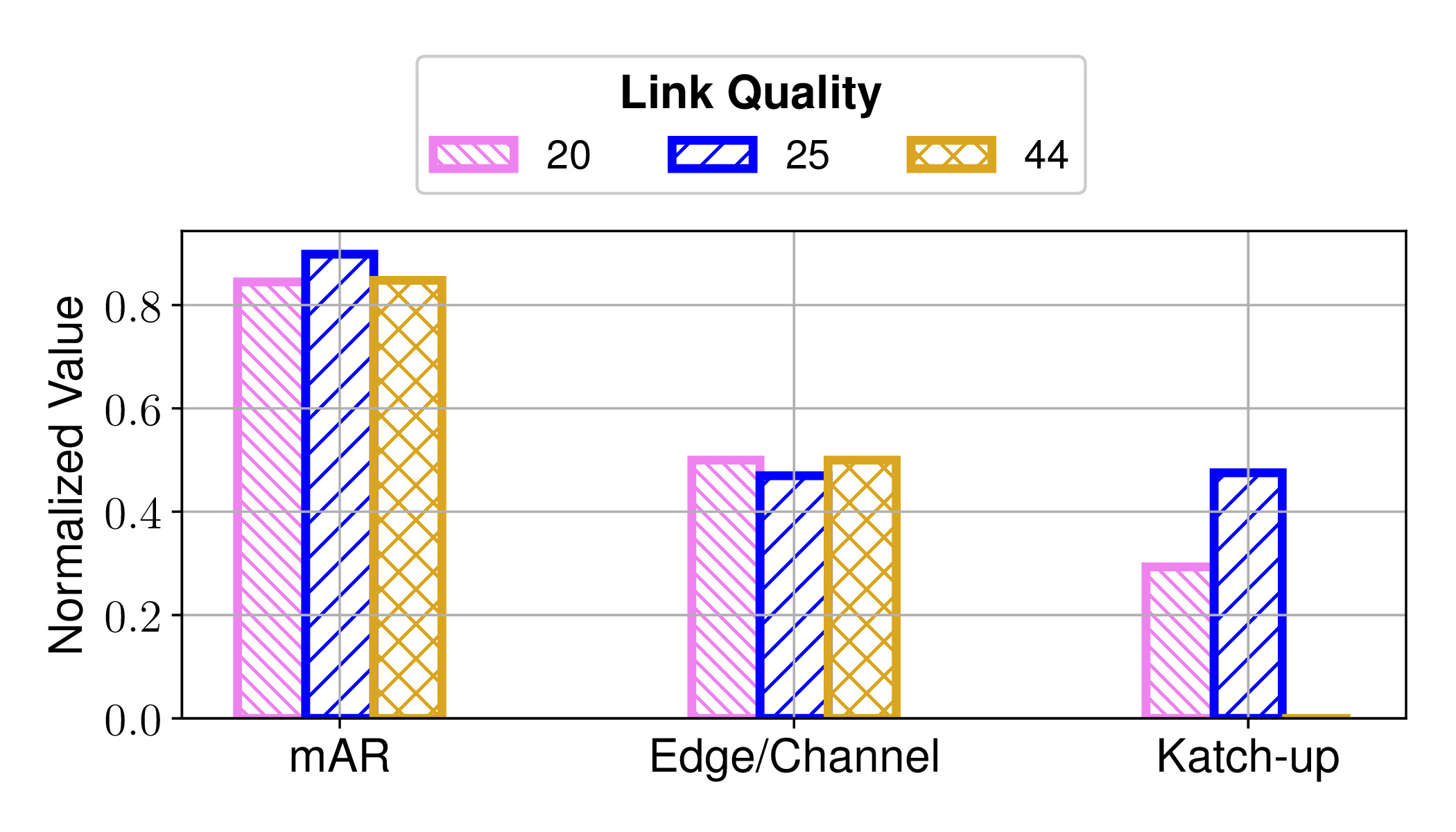}
         \caption{}
         \label{fig:linkqual}
     \end{subfigure}
     \begin{subfigure}[t]{0.33\textwidth}
         \centering
         \includegraphics[width=\textwidth,trim={5mm 5mm 5mm 5mm} ,clip]{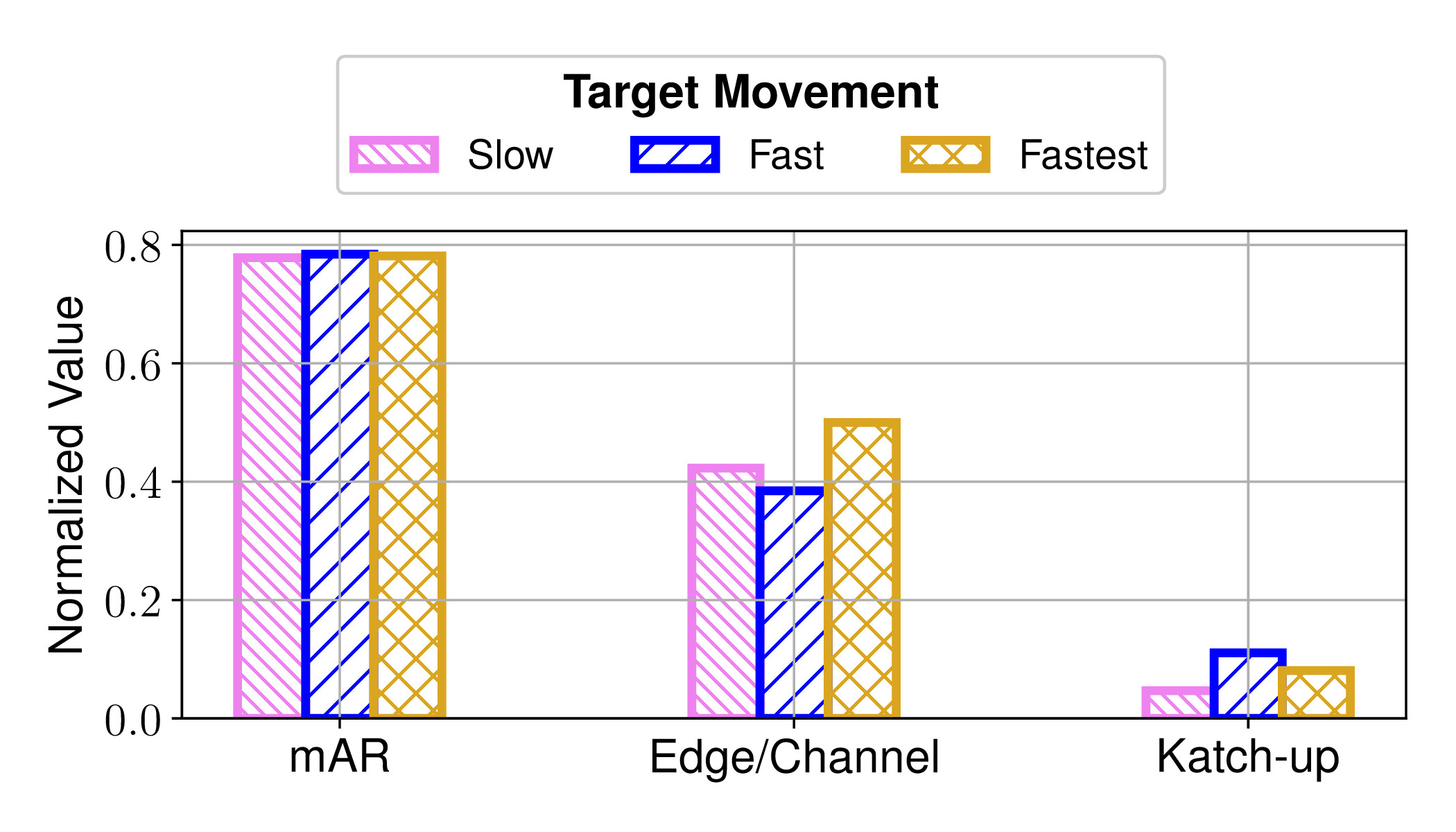}
         \caption{}
         \label{fig:mobility}
     \end{subfigure}
     \caption{(a) mAR performance and resource usage of \FW against fixed strategies. (b) mAR performance and resource usage of \FW in different link conditions. (c) mAR performance and resource usage of \FW for video sections with different target mobility.\vspace{-4mm}}
\end{figure*}


We now analyze the trends using real-world hardware and channel capacity configurations. 
Fig.~\ref{fig:prelim4} depicts the \gls{mar} as a function of the EfficienDet Model for low and high channel index of the WiFi link connecting the mobile device to the edge server and \KU on and off. In (a) and (b), we use the laptop and the server as edge server. We can see how that as the complexity of the model, and thus its execution time, increases, the configuration with \KU off suffers a degradation of \gls{mar} despite the improved quality of detection. This is due to the sensitivity of the system to latency. Conversely, when \KU is on, the performance increases when the transitioning from EfficienDet 0 to 2, and then slightly decreases from 2 to 4. Notably, we see the \KU makes the performance less sensitive the channel quality. When the server is used instead of the laptop, the smaller execution time makes the use of larger models more advantageous in both cases. However, we see how very large models (EfficientDet 3 and 4) still suffer a performance loss when the \KU is off, while they maximize performance when the \KU is on. These results further demonstrate how this selection needs to be based on features of the system in addition to characteristics of the video itself.


\subsection{\FW Evaluation}\label{sec:smartdet_results}

We now show the performance and policy structure of the \gls{drl} agent we designed and trained. In the plots, we set $\mathbf{\alpha}=[\alpha_1, \alpha_2, \alpha_3]=[0.1,0.2,0.7]$.
Fig.~\ref{fig:comparison} depicts \gls{mar}, index of channel/edge utilization (where $0, 0.5$ and $1$ correspond to period of $15, 10, 5$), \KU usage (expressed as percentage of frames to which \KU is applied) and normalized model used (EffDet 0, 2, 4 respectively mapped to $0,0.5,1$) of \FW against $2$ fixed configurations. The first fixed policy (Policy 1) maximizes resource usage by applying \KU on all the frames and setting the period to its minimum value. The fixed second policy (Policy 2) completely deactivates \KU, and sets the period to its maximum value, thus minimizing resource usage. Both policies use EfficientDet 4 to maximize \gls{od} accuracy.
In summary, by learning to dynamically adapt local analysis and offloading parameters to the current state, \FW achieves optimally controls resource usage while maximizing mAR. With respect to Policy 1, we still increase mAR by 4\% while applying \KU to only 1/3 of the frames and doubling the period. With respect to Policy 2, we increase mAR by 20\%, while increasing \KU usage to 1/3 and channel usage by 1/3.

We now analyze the decision making of the agent. In Fig.~\ref{fig:linkqual}, we show the mAR performance and channel usage, \KU activation and used model for different Wi-Fi channel quality index. First, we observe that mAR increases with channel quality as expected, and is extremely stable from low to high channel quality. Notably, the \FW agent uses different strategies for different channel qualities, thus demonstrating how the optimal parameter configuration needs to take into account contextual variables. As expected, as the channel improves the agent activates \KU less often, due to the decreased \gls{od} latency. We note that channel/edge usage and model is non-obvious. 
Indeed, for low and high channel index the agent selects simpler -- and thus faster -- EfficientDet models, while sending more frames to the edge server for \gls{od}. This behaviour exposes some of the interdependencies between the different metrics. Notice how using higher EfficientDet requires higher \KU to be performed (more complex \gls{od} has higher transmission and computation delay), and forces the agent, who is rewarded with higher \gls{mar}, to offload less frequently.

Fig.~\ref{fig:mobility} shows the metrics as a function of the target movement. Again, we see how this parameter influences the decision made by the agent, and how the agent is capable of making \gls{mar} uniform across different classes of videos. When the video has slow target movements, the \FW agent uses less channel/edge resources, activates \KU less often and uses a more complex, and thus slow, EfficientDet model compared to portions of videos where targets move fast. This strategy privileges fast \gls{od} reference turnout accepting a reference quality degradation as tracking would not be able to otherwise follow the fast targets. The more frequent \KU activation further improves the ``freshness'' of the reference. When the targets have medium mobility, the \FW agent uses a different strategy, using more complex EfficientDet models applied to fewer frames while compensating activating more often \KU.


\section{Conclusions}

This paper focused on edge-assisted real-time \gls{ot} at mobile devices, where the edge server periodically performs \gls{od} to generate references for the tracker. In this context, the key issues are (\emph{i}) remote \gls{od} may have a large and erratic latency due to channel capacity and server limitations, and (\emph{i}) the system and characteristics of the video are time-varying. To address these issues, we made two main innovations: (\emph{i}) we proposed \KU, a tracking strategy that boosts performance while sacrificing computing load and energy at the mobile device; (\emph{ii}) a \gls{drl} agent which dynamically controls tracking and offloading parameters to adapt image analysis to time-varying characteristics of the video and system variables. Results on a real-world experimental platform demonstrate the ability of the system to provide optimal tracking performance while parsimoniously using channel and energy resources.

\footnotesize
\bibliography{biblio}
\bibliographystyle{ieeetr}

\end{document}